\documentclass[prd,preprint,floatfix,nofootinbib,preprintnumbers,superscriptaddress,showkeys]{revtex4} 
\pdfoutput=1
\usepackage[caption=false]{subfig}
\usepackage{amsmath}
\usepackage{amsfonts}
\usepackage{amssymb}
\usepackage{float}
\usepackage{color}
\usepackage{graphicx}
\usepackage{graphics}
\usepackage{hyperref}
\usepackage{adjustbox,array}
\usepackage{enumitem} 
\hypersetup{}
\usepackage[utf8x]{inputenc} 
\usepackage{multirow}
\usepackage{booktabs}
\usepackage{mathrsfs}
\usepackage{diagbox}
\usepackage[english]{babel}
\usepackage[autostyle]{csquotes}

\graphicspath{{./figures/}}

\definecolor{rosso}{cmyk}{0,1,1,0.4}
\definecolor{rossos}{cmyk}{0,1,1,0.55}
\definecolor{rossoc}{cmyk}{0,1,1,0.2}
\definecolor{blu}{cmyk}{1,1,0,0.3}
\definecolor{blus}{cmyk}{1,1,0,0.6}
\definecolor{bluc}{cmyk}{1,1,0,0.1}
\definecolor{verde}{cmyk}{0.92,0,0.59,0.25}
\definecolor{verdec}{cmyk}{0.92,0,0.59,0.15}
\definecolor{verdes}{cmyk}{0.92,0,0.59,0.4}

\AtBeginDocument{
\heavyrulewidth=.08em
\lightrulewidth=.05em
\cmidrulewidth=.03em
\belowrulesep=.65ex
\belowbottomsep=0pt
\aboverulesep=.4ex
\abovetopsep=0pt
\cmidrulesep=\doublerulesep
\cmidrulekern=.5em
\defaultaddspace=.5em
}

\hypersetup{
 colorlinks=true,
 linkcolor=verdec,
 urlcolor=verdec,
 citecolor=verdec
}

\newcommand{\gsim}{\gtrsim}
\newcommand{\lsim}{\lesssim}

\newcommand{\lf}{\left(}
\newcommand{\ri}{\right)}

\newcommand{\nn}{\nonumber}

\newcommand{\sqt}{\sqrt{2}}
\newcommand{\hf}{h_{\rm f}}
\newcommand{\rr}{{\gamma\gamma}}
\newcommand{\drr}{J_{\gamma\gamma}}

\renewcommand{\lg}{\mathscr{L}} 

\newcommand{\mco}{\mathcal{O}}

\newcommand{\mcs}{\mathcal{S}}

\newcommand{\br}{\text{Br}}
\newcommand{\hc}{{\rm H.c.}}

\newcommand{\tot}{{\rm tot}}
\newcommand{\cut}{{\rm cut}}
\newcommand{\final}{{\rm final}}

\newcommand{\evfeat}{\mathbf{v}_{\rm event}}

\newcommand{\lumtot}{\mathcal{L}_{\rm tot}}

\newcommand{\fb}{{\;{\rm fb}}}

\newcommand{\iab}{{\;{\rm ab}^{-1}}}

\newcommand{\gev}{{\;{\rm GeV}}}
\newcommand{\tev}{{\;{\rm TeV}}}

\newcommand{\nsg}{N_{\rm S}}
\newcommand{\nbg}{N_{\rm B}}
\newcommand{\dbg}{\delta_{\rm B}}
\newcommand{\Dbg}{\Delta_{\rm B}}

\newcommand{\xc}{x_{\rm cut}}

\newcommand{\beq}{\begin{equation}}
\newcommand{\eeq}{\end{equation}}
\newcommand{\bea}{\begin{eqnarray}}
\newcommand{\eea}{\end{eqnarray}}
\newcommand{\barr}{\begin{array}}
\newcommand{\earr}{\end{array}}
\newcommand{\bc}{\begin{center}}
\newcommand{\ec}{\end{center}}
\newcommand{\bit}{\begin{itemize}}
\newcommand{\eit}{\end{itemize}}
\newcommand{\ben}{\begin{enumerate}}
\newcommand{\een}{\end{enumerate}}

\newcommand{\al}{\alpha}

\newcommand{\Dt}{\Delta}

\newcommand{\sg}{\sigma}

\newcommand{\kp}{\kappa}

\newcommand{\gm}{\gamma}

\newcommand{\lmc}{\Lambda_{\rm cut}}

\newcommand{\hsm}{{h_{\rm SM}}}
\newcommand{\ch}{H^\pm}

\newcommand{\wpm}{W^\pm}

\newcommand{\mch}{M_{H^\pm}}

\newcommand{\ma}{M_{A}}
\newcommand{\mach}{M_{A/H^\pm}}
\newcommand{\mhf}{m_{h_{\rm f}}}
\newcommand{\mrr}{{m_{\gamma\gamma}}}
\newcommand{\mtch}{M_T^{H^\pm}}

\newcommand{\ca}{c_\alpha}

\newcommand{\tb}{t_\beta}

\newcommand{\cba}{c_{\beta-\alpha}}
\newcommand{\sba}{s_{\beta-\alpha}}

\newcommand{\ee}      {{e^+ e^-}}

\newcommand{\ttop}      {{t\bar{t}}}

\newcommand{\bb}      {{b \bar{b}}}
\newcommand{\ww}      {{W^+ W^-}}

\newcommand{\met}      {{E_T^{\rm miss}}}
\newcommand{\vmet}      {{\vec{E}_T^{\rm miss}}}
\newcommand{\pt}      {p_T}

\definecolor{mint}{rgb}{0.24, 0.71, 0.54}

\begin{document}

\title{\color{verdes} Probing Light Fermiophobic Higgs Boson 
via diphoton jets \\ at the HL-LHC} 
\author{Daohan Wang}
\email{wdh9508@gmail.com}
\address{Department of Physics, Konkuk University, Seoul 05029, Republic of Korea}
\author{Jin-Hwan Cho}
\email{chof@nims.re.kr}
\address{National Institute for Mathematical Sciences, Daejeon 34047, Republic of Korea} 
\author{Jinheung Kim}
\email{jinheung.kim1216@gmail.com}
\address{Department of Physics, Konkuk University, Seoul 05029, Republic of Korea} 
\author{Soojin Lee}
\email{soojinlee957@gmail.com}
\address{Department of Physics, Konkuk University, Seoul 05029, Republic of Korea}
\author{Prasenjit Sanyal}
\email{prasenjit.sanyal01@gmail.com}
\address{Department of Physics, Konkuk University, Seoul 05029, Republic of Korea}
\author{Jeonghyeon Song}
\email{jhsong@konkuk.ac.kr}
\address{Department of Physics, Konkuk University, Seoul 05029, Republic of Korea}

\begin{abstract} 
In this study, we explore the phenomenological signatures associated with a light fermiophobic Higgs boson, $h_{\rm f}$, within the type-I two-Higgs-doublet model at the HL-LHC. Our meticulous parameter scan illuminates an intriguing mass range for $m_{h_{\rm f}}$, spanning $[1,10]{\;{\rm GeV}}$. This mass range owes its viability to substantial parameter points, largely due to the inherent challenges of detecting the soft decay products of $h_{\rm f}$ at contemporary high-energy colliders. Given that this light $h_{\rm f}$ ensures $\text{Br}(h_{\rm f}\to\gamma\gamma)\simeq 1$, $\text{Br}(H^\pm \to h_{\rm f} \wpm)\simeq 1$, and $M_{H^\pm}\lesssim 330{\;{\rm GeV}}$, we propose a golden discovery channel: $pp\to h_{\rm f}H^\pm\to \gamma\gamma\gamma\gamma \,\ell^\pm\nu$, where $\ell^\pm$ includes  $e^\pm$ and $\mu^\pm$.  However, a significant obstacle arises as the two photons from the $h_{\rm f}$ decay mostly merge into a single jet due to their proximity within $\Delta R<0.4$. This results in a final state characterized by two jets, rather than four isolated photons, thus intensifying the QCD backgrounds. To tackle this, we devise a strategy within \textsc{Delphes} to identify jets with two leading subparticles as photons, termed diphoton jets. Our thorough  detector-level simulations across 18 benchmark points predominantly show signal significances exceeding the $5\sigma$ threshold at an integrated luminosity of $3{\;{\rm ab}^{-1}}$. Furthermore, our approach facilitates accurate mass reconstructions for both $m_{h_{\rm f}}$ and $M_{H^\pm}$. Notably, in the intricate scenarios with  heavy charged Higgs bosons, our application of machine learning techniques provides a significant boost in significance.
\end{abstract}

\vspace{1cm}
\keywords{Higgs Physics, Beyond the Standard Model, Machine Learning}

\maketitle
\tableofcontents
\flushbottom 
\section{Introduction}

The discovery of the Higgs boson with a mass of $125\gev$ at the LHC~\cite{ATLAS:2012yve,CMS:2012qbp} 
was a pivotal moment in validating the standard model (SM). 
Beyond this foundational achievement, the Higgs boson holds an unparalleled position, 
serving as a potential portal to probe theories of particle physics beyond the SM (BSM). 
This perspective emerges from numerous unresolved fundamental questions, 
such as the nature of dark matter~\cite{Navarro:1995iw,Bertone:2004pz}, 
neutrino masses, the metastability of the SM vacuum~\cite{Degrassi:2012ry}, and the naturalness problem~\cite{Dimopoulos:1995mi,Chan:1997bi,Craig:2015pha}, all of which have deep ties to the Higgs sector. 
Therefore, postulating an extended Higgs sector is both logical and compelling.
However, despite great efforts, 
current explorations of the Higgs sector have not identified any significant deviations from the predictions of the SM: 
the properties of the observed Higgs boson align perfectly with SM expectations, 
and direct searches for additional scalar bosons have so far yielded no new findings. 
Nonetheless, the unwavering pursuit of BSM theories persists. 
One promising avenue is to probe scenarios where potential discovery channels for new Higgs bosons may have been overlooked.

A charming example is a light fermiophobic Higgs boson, $\hf$, with a mass below 125 GeV
in the type-I two-Higgs-doublet model (2HDM)~\cite{Akeroyd:1995hg,Akeroyd:1998ui,Akeroyd:1998dt,Barroso:1999bf,Brucher:1999tx,Akeroyd:2003bt,Akeroyd:2003xi,Akeroyd:2007yh,Arhrib:2008pw,Gabrielli:2012yz,Berger:2012sy,Gabrielli:2012hd,Cardenas:2012bg,Ilisie:2014hea,Delgado:2016arn,Mondal:2021bxa,Bahl:2021str,Kim:2022hvh,Kim:2023lxc,Mondal:2023wib,Sanyal:2023pfs}. 
This light mass is rationalized in the inverted Higgs scenario~\cite{Bernon:2015wef,Chang:2015goa,Song:2019aav,Jueid:2021avn,Cheung:2022ndq,Lee:2022gyf}, 
where the heavier $CP$-even Higgs boson is the observed one.
The fermiophobic nature of $\hf$ stems from the condition $\al=\pi/2$ in type-I,\footnote{Here, 
$\al$ denotes the mixing angle between the two $CP$-even Higgs bosons in the 2HDM.}
where all Yukawa couplings of $\hf$ are proportional to $\cos\al$. 
At the LHC, the production of $\hf$ is straightforward, primarily through the $pp \to W^{* } \to\hf\ch$ channel. 
Given the dominant decay modes $\hf\to\rr$ and $\ch\to \hf \wpm/\tau\nu$, 
several studies have explored new signatures such as $4\gm+V$\cite{Akeroyd:2003bt,Akeroyd:2003xi,Akeroyd:2005pr,Delgado:2016arn,Arhrib:2017wmo}, 
$4\gm+VV'$\cite{Kim:2022nmm}, and $\tau^\pm\nu \gamma\gamma$~\cite{Kim:2023lxc}.

Yet, there remains an unexplored territory for the light $\hf$ within the mass range $\mhf \in [1,10]\gev$. 
Delving into this range is essential, as it encompasses numerous parameter points 
that meet theoretical prerequisites, experimental constraints, and a cutoff scale surpassing $10\tev$. 
However, its signals at the LHC remain elusive to traditional search methodologies.
This is primarily because the two photons from the $\hf$ decay are highly collimated to merge into a single jet, not manifesting as two isolated photons.
Huge backgrounds from QCD jets should obscure the $\hf$ signals.

To tackle this challenge, 
we propose investigating the subparticles within the jet 
using EFlow objects in the \textsc{Delphes} framework~\cite{deFavereau:2013fsa}.
This novel methodology allows us to distinguish between QCD jets and signal jets housing two leading subparticles as photons, termed \enquote{diphoton jets}.
Although diphoton jets have been studied in the context of axion-like particles~\cite{Mimasu:2014nea,Bauer:2017nlg,Knapen:2021elo,Wang:2021uyb,Ren:2021prq},
no research has been conducted regarding the light fermiophobic Higgs boson.
Our study addresses this gap for the first time.

Drawing from insights on diphoton jet studies, we will execute a meticulous simulation at the detector level for the signal-to-background analysis, spanning 18 benchmark points to comprehensively represent the viable parameter space.
In the cut-based analysis, we will devise a strategy aimed at maximizing significances.
Moreover, we will illustrate the potential for accurately reconstructing the masses of $\mhf$ and $\mch$.
For challenging scenarios involving heavy charged Higgs bosons,
we will turn to machine learning techniques~\cite{Larkoski:2017jix,Guest:2018yhq,Albertsson:2018maf,Radovic:2018dip,Bourilkov:2019yoi, Feickert:2021ajf, Shanahan:2022ifi},
specifically employing one-dimensional convolutional neural networks (CNN)~\cite{CMS:2019gpd}.
The improvements achieved through this approach mark significant contributions to the topic.

The structure of this paper is outlined as follows.
In Sec.~\ref{subsec:review}, we offer a concise review of our model.
Section \ref{subsec:scan} details the scanning methodology used to determine the viable parameter space. 
We also explore the defining characteristics of these allowed parameter points, 
emphasizing the branching ratios of the BSM Higgs bosons.
In Sec.~\ref{subsec:signal}, the unique feature that the two photons from the $\hf$ decay appear as a single jet isx clarified.
Section \ref{sec:jet:substructure} is dedicated to discussing the phenomenologies of the subparticles within the diphoton jet. 
We also provide a new method to subtract  the significant pileups anticipated at the HL-LHC.
In Sec.~\ref{sec:cut-based}, we direct our focus towards the signal-to-background analysis in a cut-based approach.
Section \ref{sec:mass:reconstruction} sees us undertaking the task of mass reconstruction for both $\mhf$ and $\mch$. 
For the challenging cases involving heavy charged Higgs bosons, machine learning techniques come into play. These are detailed in Sec.~\ref{sec:CNN}.
Finally, our conclusions are presented in Sec.~\ref{sec:conclusions}.

\section{Fermiophobic type-I with very light $\hf$}

\subsection{Review of the fermiophobic type-I}
\label{subsec:review}
Let us briefly review the type-I 2HDM with a light fermiophobic Higgs boson. 
The 2HDM introduces two $SU(2)_L$ complex scalar doublet fields with hypercharge $Y=1$~\cite{Branco:2011iw}:
\bea
\label{eq:phi:fields}
\Phi_i = \left( \begin{array}{c} w_i^+ \\[3pt]
\dfrac{v_i +  \rho_i + i \eta_i }{ \sqrt{2}}
\end{array} \right) \quad \text{for $i=1,2$.} 
\eea
Here, $v_{1}$ and $v_2$ denote the vacuum expectation values of $\Phi_{1}$ and $\Phi_2$, respectively,
defining $\tan\beta  = v_2/v_1$.
The electroweak symmetry is spontaneously broken by $v =\sqrt{v_1^2+v_2^2}=246\gev $.
For the sake of simplicity in notation,  we will use $s_x =\sin x$, $c_x = \cos x$, and $t_x = \tan x$ in what follows.

In order to prevent flavor changing neutral currents (FCNCs) at the tree level,
a discrete $Z_2$ symmetry is imposed,
under which $\Phi_1 \to \Phi_1$ and $\Phi_2 \to -\Phi_2$~\cite{Glashow:1976nt,Paschos:1976ay}.
Assuming $CP$-invariance and softly broken $Z_2$ symmetry, the scalar potential is written as
\bea
\label{eq:VH}
\begin{split}
V_\Phi &=  m^2 _{11} \Phi^\dagger_1 \Phi_1 + m^2 _{22} \Phi^\dagger _2 \Phi_2
-m^2 _{12} ( \Phi^\dagger_1 \Phi_2 + \hc) + \frac{\lambda_1}{2} (\Phi^\dagger _1 \Phi_1)^2 
+ \frac{\lambda_2 }{2}(\Phi^\dagger _2 \Phi_2 )^2 \\
&\phantom{=}\quad
+ \lambda_3 (\Phi^\dagger _1 \Phi_1) (\Phi^\dagger _2 \Phi_2) 
+ \lambda_4 (\Phi^\dagger_1 \Phi_2 ) (\Phi^\dagger _2 \Phi_1) 
+ \frac{\lambda_5}{2} 
\left[
(\Phi^\dagger _1 \Phi_2 )^2 +  \hc
\right].
\end{split}
\eea

Within this framework, five physical Higgs bosons emerge:  the lighter $CP$-even scalar $h$,
the heavier $CP$-even scalar $H$, the $CP$-odd pseudoscalar $A$,
and a pair of charged Higgs bosons $H^\pm$.
These physical Higgs bosons are related with the weak eigenstates in \autoref{eq:phi:fields}
through two mixing angles, namely $\alpha$ and $\beta$~\cite{Song:2019aav}.
The SM Higgs boson $\hsm$ is a linear combination of $h$ and $H$,
expressed as $\hsm = \sba h + \cba H$.
Since the Higgs boson observed at the LHC has shown remarkable alignment with the predicted behavior of $\hsm$~\cite{ATLAS:2020fcp,ATLAS:2020bhl,CMS:2020zge,ATLAS:2021nsx,CMS:2021gxc,ATLAS:2020syy,ATLAS:2021upe,ATLAS:2020pvn,CMS:2021ugl,ATLAS:2020wny,ATLAS:2020rej,ATLAS:2020qdt,ATLAS:2020fzp,CMS:2020xwi,ATLAS:2021zwx},
we have two plausible scenarios, the normal scenario where $h \simeq \hsm$ and the inverted scenario where $H \simeq \hsm$.
To accommodate a light fermiophobic Higgs boson, 
we focus on type-I within the inverted Higgs scenario. 
In type-I, every Yukawa coupling associated with $h$ is proportional to $\ca$.
Therefore, by merely setting $\alpha=\pi/2$, 
$h$ acquires fermiophobic characteristics, which endure even when loop corrections are considered~\cite{Barroso:1999bf,Brucher:1999tx}.
For brevity in subsequent discussions,
we will denote the type-I 2HDM with $\al=\pi/2$ in the inverted Higgs scenario as the fermiophobic type-I 
and the lighter $CP$-even Higgs boson with $\al=\pi/2$ as $\hf$. 

The Yukawa couplings of the SM fermions are parametrized by
\begin{align*}
\lg^{\rm Yuk} =
&- \sum_f 
\lf 
\frac{m_f}{v} \xi^h_f \bar{f} f \hf + \frac{m_f}{v} \kp_f^H \bar{f} f H
-i \frac{m_f}{v} \xi_f^A \bar{f} \gm_5 f A
\ri
\\
&-
\left\{
\dfrac{\sqrt{2}}{v } \overline{t}
\left(m_t \xi^A_t P_- +  m_b \xi^A_b P_+ \right)b  H^+
+\dfrac{\sqt m_\tau}{v}\xi^A_\tau \,\overline{\nu}_\tau P_+ \tau H^+
+\hc
\right\},
\end{align*}
where $P_{\pm} = (1\pm \gm^5)/2$.
In the fermiophobic type-I, the Yukawa coupling modifiers are given by
\begin{align}
\label{eq:Yukawa:couplings}
\xi^{\hf}_f&=0 ,
\quad
\kp^H_f = \frac{\sqrt{1+\tb^2}}{\tb} ,
\quad
\xi^A_t = -\xi^A_b = -\xi^A_\tau = \frac{1}{\tb}.
\end{align}
To be consistent with the current best-fit results for the Peskin-Takeuchi oblique parameters~\cite{Peskin:1991sw}, 
an additional assumption is introduced: $\ma=\mch\equiv \mach$.
In summary, the complete set of model parameters includes:
\bea
\label{eq:model:parameters}
\{ \mhf, \; \mach,\; m_{12}^2,\; \tb \}.
\eea

\subsection{Viable parameter space for very light $\hf$}
\label{subsec:scan}

In the quest to discover the light $\hf$ at the LHC, 
our preliminary task involves a systematic scan of the parameter space 
to identify viable candidates that comply with both theoretical requirements and experimental constraints. 
Our scan encompasses the following ranges:
\begin{align}
\label{eq:scan:range}
& \mhf \in [1,30]\gev, \quad
\mach \in \left[ 80, 900 \right] \gev, \\ \nn
& \tb \in \left[ 0.5,50 \right], \qquad\quad
m_{12}^2 \in \left[ 0, 20000 \right] \gev^2.
\end{align}
We consider only positive values for $m_{12}^2$ 
since preliminary scans indicate that parameter points with negative $m_{12}^2$ fail to meet the vacuum stability condition.

Within this extensive parameter space, 
we apply a cumulative series of constraints, outlined as follows:\footnote{Due to our assumption $\mch = \ma$, we disregard constraints from the Peskin-Takeuchi oblique parameters, as the new contributions from the BSM Higgs bosons become negligible~\cite{He:2001tp,Grimus:2008nb}.}
\begin{description}
\item[Step A.]Theoretical requirements and the low energy data
\renewcommand\labelenumi{(\theenumi)}	
	\ben
	\item We use the public code \textsc{2HDMC} to ensure the bounded-from-below condition for the Higgs potential~\cite{Ivanov:2006yq}, tree-level unitarity of scalar-scalar scatterings~\cite{Branco:2011iw,Arhrib:2000is}, and perturbativity of the Higgs quartic couplings~\cite{Chang:2015goa}.
	Additionally, the vacuum stability condition is enforced~\cite{Ivanov:2008cxa,Barroso:2012mj,Barroso:2013awa}.
	\item 
	We demand alignment with the FCNC data, particularly emphasizing the inclusive $B$-meson decay measurements into $X_s \gamma$ at the 95\% C.L.~\cite{Arbey:2017gmh,Sanyal:2019xcp,Misiak:2017bgg}.
	\item 
	We require the cutoff scale $\lmc$ to exceed $10\tev$. To determine this, we run the model parameters under the renormalization group equations using the public \textsc{2HDME} code~\cite{Oredsson:2018yho,Oredsson:2018vio}. The cutoff scale is defined by the energy scale at which any of the three conditions—tree-level unitarity, perturbativity, or vacuum stability—is violated~\cite{Kim:2023lxc}.
	\een
\item[Step B.]   High energy collider data
\renewcommand\labelenumi{(\theenumi)}
	\ben
	\item We examine direct search constraints from LEP, Tevatron, and LHC experiments, 
	excluding parameter points with a cross section above the observed $2\sg$ band.
	We used the public code \textsc{HiggsBounds}-v5.10.2~\cite{Bechtle:2013wla}.
	\item We assess alignment with Higgs precision data utilizing \textsc{HiggsSignals}-v2.6.2~\cite{Bechtle:2020uwn}. We mandate that the cross section of a parameter point lies within $2\sigma$ confidence levels in relation to the model's optimal fit point.
	\item We consider additional measurements sensitive to the light fermiophobic Higgs boson. This includes  $\ee\to \hf(\to\rr)Z$, $\ee \to \hf(\to\rr)A(\to \bb/\hf Z)$~\cite{DELPHI:2003hpv}, $p\bar{p}\to \hf \ch (\to \hf \wpm) \to 4\gamma X$~\cite{CDF:2016ybe}, and $pp \to H \to \hf\hf \to 4\gamma$~\cite{CMS:2021bvh}. Parameter points yielding a cross section above the $2\sigma$ bound are excluded.
	\een
\end{description}

Let us begin by examining the survival rates after each constraint is applied. 
We use the parameter points that satisfy Step A(1) as our reference dataset, from which all subsequent survival rates are calculated. 
Upon implementing the FCNC constraint in Step A(2), a respectable 73.3\% of points persist. 
The enforcement of $\lmc>10\tev$ in Step A(3) further refines our pool, leaving 26.6\% of points standing. 
Progressing to Step B(1), our selection tightens, whittling down to a mere 2.03\%. 
Upon assimilation of the Higgs precision data in Step B(2), around 1.94\% survive. 
Ultimately, after accounting for Step B(3), 1.38\% of the parameter points from A(1) endure.

\begin{figure}[!t]
\centering
\includegraphics[width=\textwidth]{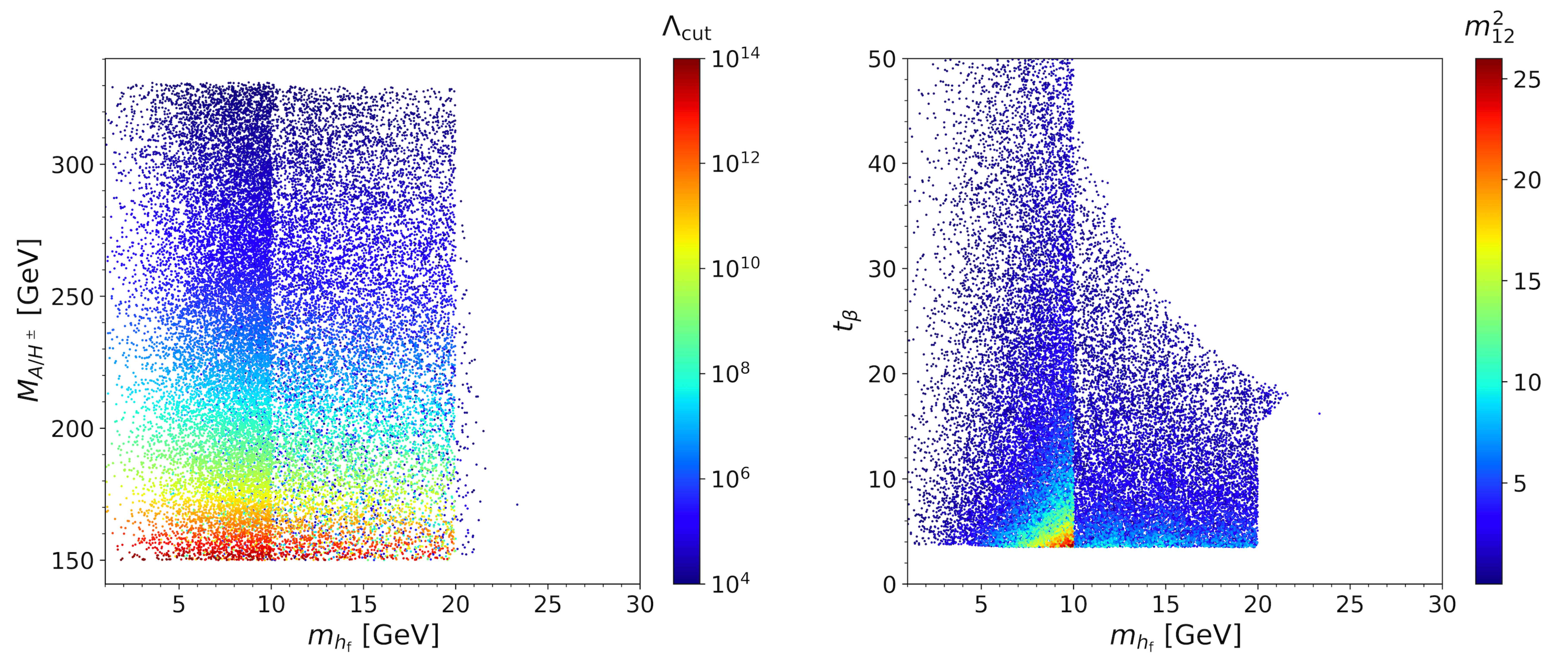}
\vspace{-0.7cm}
\caption{$\mch$ versus $\mhf$ with a color-code of $\lmc$ in GeV (left panel), 
and $\tb$ versus $\mhf$ with a color-code of $m_{12}^2$ in units of ${\rm GeV}^2$ (right panel). 
All depicted parameter points satisfy the complete set of theoretical and experimental constraints.
The parameter points are ordered  by ascending values of $\lmc$ in the left panel and $m_{12}^2$ in the right panel. }
\label{fig-characteristics}
\end{figure}

Now we investigate the characteristics of the parameter points satisfying all imposed constraints.
In \autoref{fig-characteristics}, we present   $\mch$ versus $\mhf$ with the color code of $\lmc$ (left panel), 
and $\tb$ versus $\mhf$ with the color code of $m_{12}^2$ (right panel). 
For visualization clarity, 
we have ordered the parameter points by ascending values of $\lmc$ in the left panel and $m_{12}^2$ in the right panel. 
This stacking method ensures that points with lower $\lmc$ (or $m_{12}^2$) are positioned underneath~\cite{Kang:2022mdy}.

Turning to the $\mach$ versus $\mhf$ plot, we notice several distinct features. 
First, the density of viable parameter points varies noticeably with the $\mhf$ value. 
Specifically, the number of viable parameter points per unit mass for the intervals $[1,10]\gev$, $[10,20]\gev$, and $[20,30]\gev$ has a ratio of 1\,:\,0.71\,:\,0.0058.
These significant variations arise from the following direct search constraints:
\bit
\item The measurement of $pp \to \hsm \to \hf\hf \to 4\gamma$ by the ATLAS Collaboration
significantly constrains the parameter space for $\mhf \in[10,30]\gev$~\cite{Aad:2015bua}.
\item The examination of $\ee \to \hf Z \to \rr Z$ by the ALEPH Collaboration eliminates nearly all parameter points in $\mhf \in [20,30]\gev$~\cite{ALEPH:2002gcw}.
\eit
Considering the markedly higher survival percentages, the  mass range of $\mhf\in[1,10]\gev$ warrants thorough investigation,\footnote{A high survival percentage alone does not inherently validate any model parameter, since nature chooses one parameter point. But prioritizing parameter regions with a higher likelihood is a prudent strategy.}
 an endeavor not yet undertaken in existing literature.
 The second notable feature is 
the presence of the upper bound on $\mach$, approximately at $330\gev$.
This upper bound exhibits a tendency to decrease as $\lmc$ increases:
when $\lmc > 100\tev$, the upper threshold reduces\footnote{The inverse is not necessarily true:
a smaller $\mach$ does not automatically imply a larger $\lmc$.
Note that the blue points are positioned below the red ones.} to $\mach \lsim 280\gev$.
These features hold promising implications for the HL-LHC, 
where the center-of-mass energy of $14\tev$ offers a favorable environment for producing $\ch$.

In the $\tb$ versus $\mhf$ plot, three significant features stand out. 
First, lower bounds on $\tb$ emerge, characterized by $\tb \gsim 4$.
This happens because the Yukawa couplings of the BSM Higgs bosons increase as $\tb$ decreases, 
as illustrated in \autoref{eq:Yukawa:couplings}.
The second salient feature is an evident  transition at $\mhf \simeq 10\gev$.
Beneath this threshold, the distribution of permissible parameter points uniformly spans the $\tb\in [4,50]$ range.
For $\mhf>10\gev$, however, there is an upper limit on $\tb$, progressively declining as $\mhf$ increases.
This transition around $\mhf=10\gev$ stems from the notably light mass of $\hf$, leading to decay products in high-energy colliders that are challenging to discern.
Finally, the $m_{12}^2$ distribution primarily leans towards the lower end, peaking around $26\gev^2$.
This small $m_{12}^2$ hints the approximate preservation of $Z_2$ parity in the fermiophobic type-I, 
because only the $m_{12}^2$ term 
breaks $Z_2$ parity.

Given these characteristics of the fermiophobic type-I model,
we concentrate on the following mass range for $\hf$:
\beq
\label{eq:very:light:hf}
\mhf \in [1,10]\gev.
\eeq
In subsequent discussions and investigations, we will refer to $\hf$ within this mass range as a \enquote{very light} $\hf$.

\begin{figure}[!t]
\centering
\includegraphics[width=0.95\textwidth]{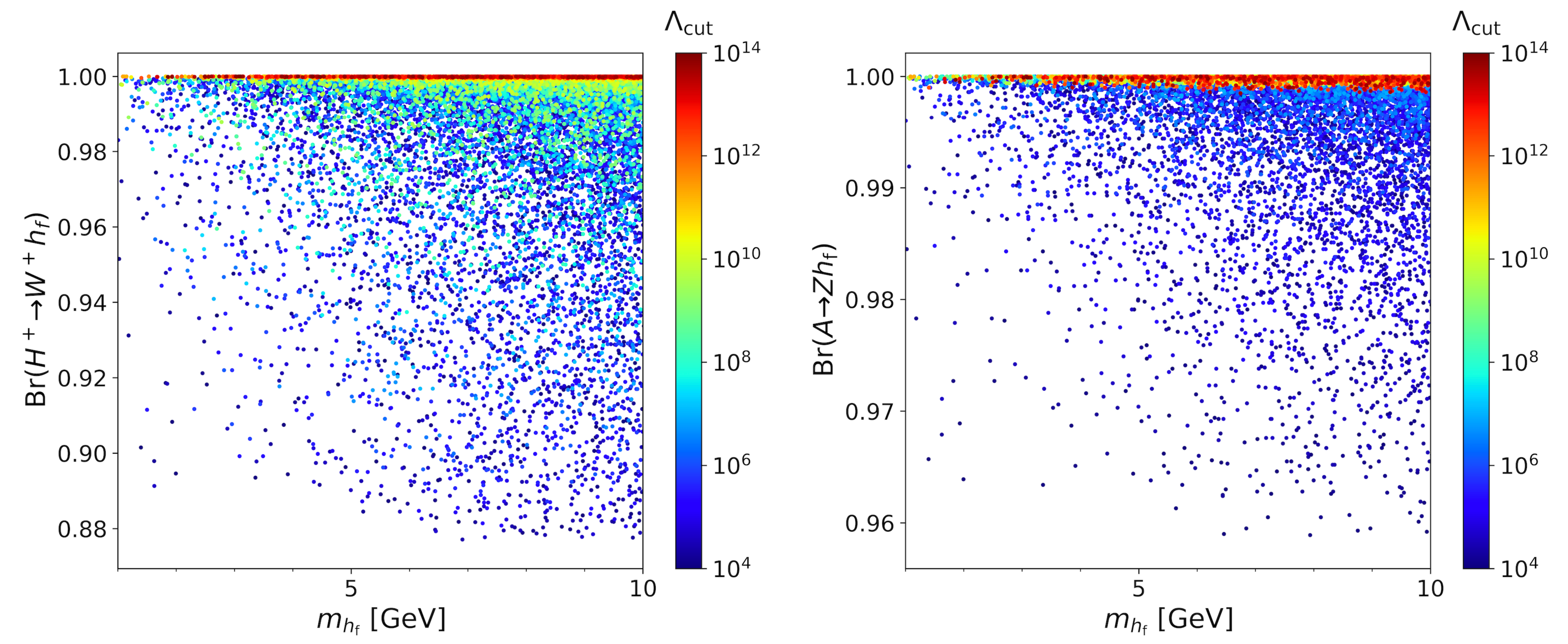}
\caption{
$\br(\ch\to\hf\wpm)$ versus $\mhf$ (left panel)
and $\br(A\to\hf Z)$ versus $\mhf$ (right panel).
The color code denotes the cutoff scale $\lmc$ in units of GeV.}
\label{fig-BRs}
\end{figure}

Given the distinct characteristics of the fermiophobic type-I model, 
our attention is directed towards the discovery potential of the the HL-LHC for the very light $\hf$. 
Central to this are its decay modes and production channels. 
The decay pattern for this particle is unambiguous, with $\br(\hf\to\rr)\simeq 100\%$. 
Its primary production mechanisms at the LHC occur in association with other BSM Higgs bosons, specifically $pp\to W^* \to \hf\ch$ and $pp \to Z^*\to \hf A$~\cite{Kim:2022nmm,Kim:2023lxc}. As a result, the final states arising from these production avenues are intrinsically tied to the decay patterns of $\ch$ and $A$.

\begin{figure}[!h]
\centering
\includegraphics[width=0.65\textwidth]{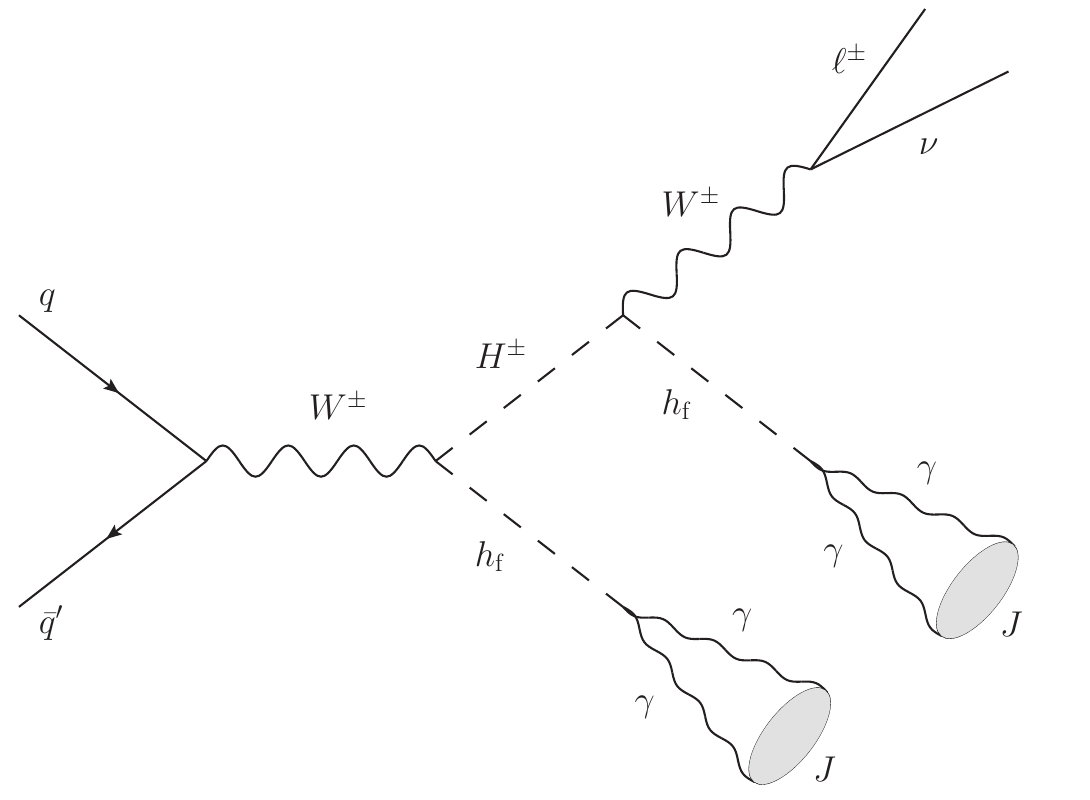}
\caption{
Feynman diagram for the signal process $p p \to W^* \to \hf \ch (\to \hf\wpm)\to \rr+ \rr + \ell^\pm \nu$. 
As the two photons from the $\hf$ decay are highly collimated, they are probed as a single jet $J$.}
\label{fig-Feynman}
\end{figure}

In \autoref{fig-BRs}, we depict $\br(\ch\to\wpm\hf)$ versus $\mhf$ (left panel) and $\br(A \to Z\hf)$ versus $\mhf$ (right panel) 
across all the viable parameter points, with the color codes signifying $\lmc$ values in GeV. 
Notably, $\ch\to\wpm\hf$ and $A\to\hf Z$ surface as the predominant decay channels, 
with $\br(\ch\to\wpm\hf)$ and $\br(A \to Z\hf)$ surpassing 88\% and 96\%, respectively. 
A high cutoff scale, such as $\lmc \sim 10^{14}\gev$, results in nearly 100\% branching ratios for both $\ch\to\hf\wpm$ and $A\to\hf Z$.
Hence, two primary candidates for discovery channels present themselves: 
$pp \to \hf \ch(\to\hf \wpm)$ and $pp \to \hf A(\to \hf Z)$.
Considering the dominant charged-current production and 
the larger branching ratio of the leptonic decays of $\wpm$ compared to $Z$,  
we propose the following as the golden channel to probe the very light $\hf$:
\bea
\label{eq:signal:process}
p p \to W^* \to \hf \ch (\to \hf\wpm)\to \rr+ \rr + \ell^\pm \met,
\eea
where $\ell^\pm=e^\pm, \mu^\pm$.
In our comprehensive analysis, we also incorporate the decay mode $\wpm\to \tau^\pm\nu$,
which is subsequently followed by $\tau^\pm \to \ell^\pm \nu\nu$.
The corresponding Feynman diagram is depicted in \autoref{fig-Feynman}.

\subsection{Signature of the golden channel $pp\to\hf\ch$}
\label{subsec:signal}

Let us now present the parton-level cross section of the proposed golden channel for $\hf$.
Initially, we generated the Universal FeynRules Output (UFO)~\cite{Degrande:2011ua}
for the fermiophobic type-I
through \textsc{FeynRules}~\cite{Alloul:2013bka}.
Incorporating this UFO file into \textsc{MadGraph5-aMC@NLO}~\cite{Alwall:2011uj},
we determined the cross-sections of $pp \to \ch\hf$ at the 14 TeV LHC.
For the parton distribution function, we adopted the \textsc{NNPDF31\_lo\_as\_0118} set~\cite{NNPDF:2017mvq}.
The branching ratios of $\hf$ and $H^\pm$ were obtained from \textsc{2HDMC}~\cite{Eriksson:2009ws},
and subsequently multiplied by the cross-sections.

\begin{figure}[!t]
\centering
\includegraphics[width=0.6\textwidth]{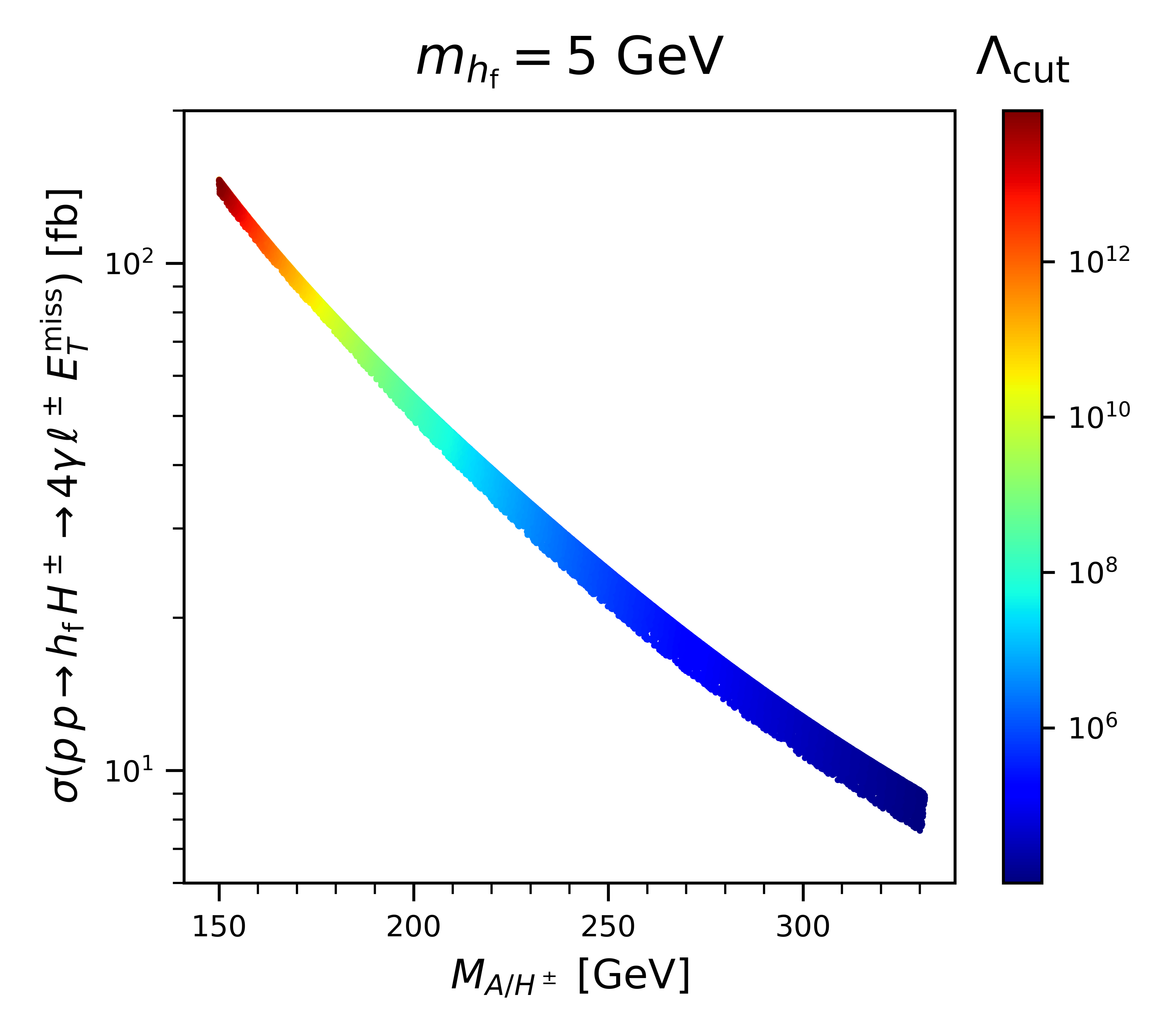}
\vspace{-0.4cm}
\caption{\label{fig:parton:xsec}
Parton-level cross sections of $pp \to \hf\ch(\to \hf \wpm) \to 4\gm \,\ell^\pm\met$
at the 14 TeV LHC, about $\mch$.
The color code represents the cutoff scale $\lmc$.
Here, we set $\mhf=5\gev$.
  }
\end{figure}

In \autoref{fig:parton:xsec}, the scatter plot shows the parton-level cross sections for $\mhf=5\gev$ 
against the charged Higgs boson mass, 
 spanning all viable parameter points.\footnote{According to our analysis,
the cross sections for cases with $\mhf=1\gev$ and $\mhf=10\gev$ align closely with those for $\mhf=5\gev$,
mostly deviating by about 1\%.}
The color code represents $\lmc$.
An expected correlation appears between the cross section and $\mch$:
as $\mch$ increases, $\sg_\tot$ decreases. 
Additionally, for a given $\mch$, the cross sections across all viable parameter points are nearly constant,
with deviations of less than 10\%.
A compelling feature is the substantial size of the signal cross section. 
Even the minimum cross section, encountered when $\mch\simeq 330\gev$, reaches a significant $\sim7\fb$.

Despite these considerable signal cross sections, 
distinguishing the signal from the background at the HL-LHC remains a challenge. 
At first glance, a final state comprised of four photons, a lepton, and missing transverse energy 
might seem to suppress major QCD backgrounds. 
But the reality is more intricate.
When the $\hf$ decays into two photons at high-energy colliders, the resulting photons are not typically isolated
because they are tightly collimated within a radius of $\Delta R < 0.4$.
Here $\Delta R$ is the angular distance, given by $\Delta R= \sqrt{(\Dt \eta)^2 + (\Dt\phi)^2}$.
Still, these photons register an energy deposit in the calorimeters, eventually being recognized and grouped as a jet.
This leads to substantial QCD backgrounds.

To better elucidate how the detector processes photons,  
let us briefly review the photon isolation criteria adopted by the \textsc{Delphes}. 
Consider a photon candidate \texttt{P}, a stable particle that deposits its energy into the electromagnetic calorimeter (ECAL), 
while leaving no trace in the tracker. 
For \texttt{P} to be recognized as a photon,
it should be sufficiently isolated from neighboring particles. 
In \textsc{Delphes},
this isolation is determined using the criterion $I(\mathtt{P})>I_{\rm min}$. 
Here, the isolation variable $I(\mathtt{P})$ is expressed as:
\bea
\label{eq:I}
I(\mathtt{P}) = \dfrac{\sum_{i\neq \mathtt{P}}^{ \Dt R<R_\gm} p_T^i }{p_T^{\mathtt{P}}},
\eea
where the numerator represents the combined transverse momenta of all particles (excluding \texttt{P}) 
that fall within a cone of radius $R_\gm$ centered around \texttt{P}.
In the \texttt{delphes\_card\_HLLHC.tcl} utilized in subsequent analysis,
the default settings are $I_{\rm min}=0.1$ and $R_\gm=0.3$.

In \textsc{Delphes}, the photon isolation is evaluated concurrently with jet clustering. 
This procedure involves clustering EflowPhoton, EflowNeutralHadrons, and EflowChargedHadrons 
according to the energy flow algorithm. 
Once this process concludes, the definitive identification for \texttt{P} is set. 
If \texttt{P} satisfies the photon isolation criteria,
it is recognized as a photon.
Conversely, if \texttt{P} fail the  criteria, it is designated as a jet. 

To demonstrate our claim that the two photons from $\hf\to\rr$ are more likely to be recognized as a single jet, 
we conducted a comprehensive detector simulation for the signal with $\mhf=5\gev$, $\mach=150\gev$, and $\br(\ch\to\hf\wpm)=\br(\hf\to\rr)=1$.
Parton showering and hadronization were integrated using \textsc{Pythia} version \texttt{8.309}~\cite{Bierlich:2022pfr}. 
We employed \textsc{Delphes} high-luminosity card \texttt{delphes\_card\_HLLHC.tcl}. 
For jet clustering, \texttt{FastJet} version 3.3.4~\cite{Cacciari:2011ma}, deploying the anti-$k_T$ algorithm~\cite{Cacciari:2008gp},
was utilized for the jet radius of $R=0.4$. 
At this stage, we opted not to consider pileup effects.

\begin{figure}[!t]
\centering
\includegraphics[width=0.8\textwidth]{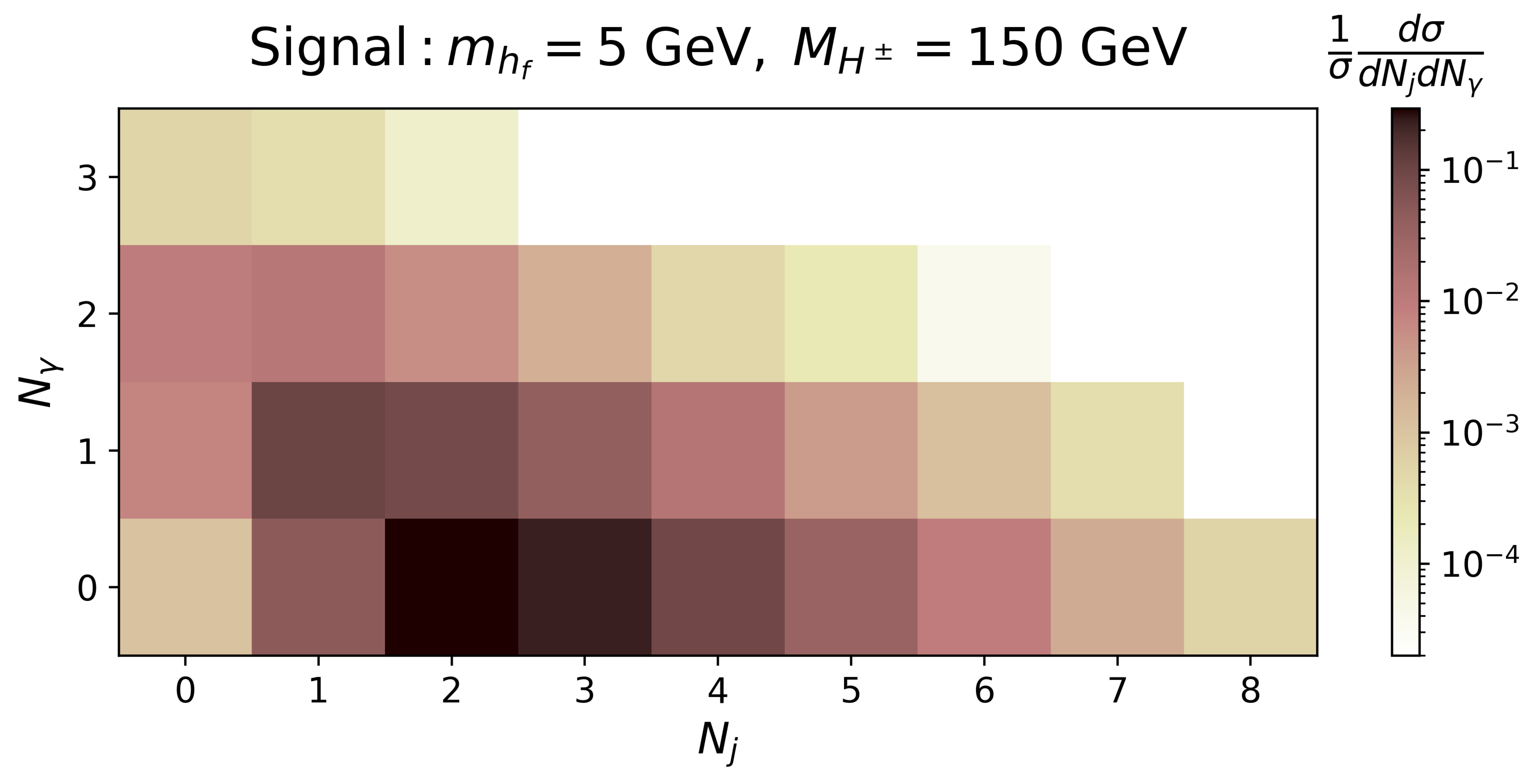}
\vspace{-0.4cm}
\caption{\label{fig-nr-nj-conventional}
The distribution of photon multiplicity versus jet multiplicity for the signal process $pp \to \hf \ch \to \rr\rr\, \ell^\pm \met$ 
following the detector simulation at the 14 TeV LHC. 
The color code indicates the normalized number of events. 
Parameters are set as $\mhf=5\gev$, $\mach=150\gev$, and $\br(\ch\to\hf\wpm)=\br(\hf\to\rr)=1$.
  }
\end{figure}

In \autoref{fig-nr-nj-conventional}, we present the photon multiplicity versus the jet multiplicity for the signal process 
at the detector level, using a color code to represent the normalized number of events.
These results are derived from $5\times 10^5$ events at the generation level. 
The findings in \autoref{fig-nr-nj-conventional} are striking.
The signal event, which includes four photons at the parton level, results in a markedly different outcome at the detector level. 
Approximately 70\% of events fall under $N_\gamma=0$, while around 26\% are categorized as $N_\gamma=1$.
Events with $N_\gamma=2$ are scarce. 
Instead, the majority of signal events manifest as two jets.

\begin{table}
  {\renewcommand{\arraystretch}{1.2} 
\begin{tabular}{|c|c|c||c|c|c|}
\hline
Background & ~Cross section [pb]~ & ~$n_{\rm gen}$ & Background & ~Cross section [pb]~ & ~$n_{\rm gen}$ \\ \hline
$\wpm(\to L^\pm\nu)jj$ & $3. 54 \times 10^3$ & ~~$5\times 10^{8}$~~  &  $\wpm Z$ & $3.16\times 10$  & ~$3\times 10^6$~  \\ \hline
$Z(\to L^+ L^-)jj$ & $2.67\times 10^2$ & $5 \times 10^7$  & $Z(\to L^+ L^-) j \gm$ & $2.09$ & $10^6$ \\ \hline
~$\ttop(\to \bb W_{L\nu} W_{jj})$~ & $ 1.23 \times 10^2$ & $1.2\times 10^7$   & ~~$ZZ$~~ & $1.18\times 10$ & $10^6$  \\ \hline
$\wpm(\to L^\pm \nu)j\gm$ & $2.53\times 10$ & $3\times 10^6$  & ~~$\wpm(\to L^\pm \nu)\rr$~~ & $3.28\times 10^{-2}$ & $10^6$ \\ \hline
~~~$\ww$~~~ & $8.22 \times 10$ & $9\times 10^6$ & $Z(\to L^+ L^-)\rr$ & $1.12\times 10^{-2}$ & $10^6$ 
\\ \hline
\end{tabular}
}
\caption{\label{table:BG:xsec} Parton-level cross sections of the backgrounds at the 14 TeV LHC, where $L^\pm$ denotes $e^\pm, \mu^\pm,$ or $\tau^\pm$.  The number of generated events, denoted as $n_{\rm gen}$, is also provided.
}
\end{table}

As the final state includes two jets, various backgrounds arise. 
We take into account a total of ten background processes: 
$\wpm (\to L^\pm\nu)jj$, $Z(\to L^+ L^-)jj$, $\ttop(\to \bb W_{L\nu} W_{jj})$, $\wpm(\to L^\pm \nu)j\gm$, $\ww$, $\wpm Z$, 
$Z(\to L^+ L^-) j \gm$, $ZZ$, $\wpm(\to L^\pm \nu)\rr$, and $Z(\to L^+ L^-)\rr$. 
Here, $L^\pm$ represents $e^\pm, \mu^\pm,$ or $\tau^\pm$. 
Given that our signal process includes either one electron or one muon,
we have incorporated the leptonic decays of $\wpm$ and $Z$ into some dominant backgrounds.

In \autoref{table:BG:xsec}, we summarize the parton-level cross sections for the ten background processes at the 14 TeV LHC, 
applying generation-level cuts of $p_T^{j}>20\gev$, 
$p_T^{L,\gm}>10\gev$, $|\eta_j|<5$, $|\eta_{L,\gm}|<2.5$, and $\Delta R_{i i'} > 0.4$
where $i$ and $i'$ include all the particles in the final state. 
Due to the considerable differences in the cross sections among these background processes,
we produce different event counts at the generation level, represented by $n_{\rm gen}$ in \autoref{table:BG:xsec}.
Notably, the background cross sections significantly exceed the signal cross section. 
If the analysis only considers collective objects like jets in the final state, 
distinguishing the signal from the backgrounds becomes almost infeasible. 
Consequently, devising a strategy targeting diphoton jets is pivotal for detecting the signal at the HL-LHC.

\section{Jet subparticles and pileups}
\label{sec:jet:substructure}

In the previous section, we illustrated that the four photons in our signal process, $pp\to \hf\hf \wpm \to 4\gm \wpm$, 
are predominantly tagged as two jets, not isolated photon entities. 
Given that these photons exist as subparticles within a jet, distinguishing this unique diphoton jet from a standard QCD jet necessitates a thorough analysis of the jet's subparticles. 
To enable this differentiation, we employ the EFlow objects within jets in the \textsc{Delphes} framework. 
These EFlow objects are divided into three categories:  EflowPhoton, EflowNeutralHadrons, and EeflowChargedHadrons, 
with each type determined by tracker and tower information.
The tracker identifies charged particles through their characteristic ionization patterns within its system, 
while tower data focus on energy deposits in the calorimeter.

To enhance our understanding, let us revisit the interactions of particles within calorimeters.
Photons, when passing through the electromagnetic calorimeter (ECAL), trigger an energy dispersion across its layers.   
Hadrons, on the other hand, deposit energy differently depending on their type. 
Neutral pions, for instance, decay promptly into a pair of photons, largely concentrating their energy within the ECAL. 
Meanwhile, stable hadrons like neutrons and charged pions predominantly channel their energy to the hadron calorimeter (HCAL). 
A notable scenario occurs with long-lived hadrons, such as Kaons and $\Lambda$ baryons.
With a decay length around 10 mm, they  interact with both the ECAL and HCAL, 
resulting in a division of energy deposit as $f_{\rm ECAL}=0.3$ and $f_{\rm HCAL}=0.7$.

Yet, when it comes to utilizing jet subparticles, the issue of pileup poses a formidable challenge.
Pileup, a byproduct of the high luminosity in hadron colliders, 
results from multiple proton-proton collisions within a single bunch crossing. 
At the HL-LHC, where roughly 200 pileup events are standard, 
discerning the diphoton jet from a QCD jet becomes intricate due to the flood of pileup-induced particles. 
Therefore, it is crucial to effectively subtract pileups in our analysis.

Several methods for pileup subtraction have been advanced, such as the jet vertex fraction~\cite{TheATLAScollaboration:2013pia}, 
charged hadron subtraction (CHS)~\cite{CMS:2014ata,Kirschenmann:2014dla}, 
the {\sc Puppi} method~\cite{Bertolini:2014bba}, and the \textsc{SoftKiller} method~\cite{Cacciari:2014gra}. 
In our exploration, we cast a special focus on CHS and \textsc{SoftKiller}. 
The CHS technique leverages the capability of the detector to determine the vertex distance of charged tracks relative to the primary vertex.
In contrast, \textsc{SoftKiller} is a fast event-level pileup subtraction tool, 
relying on a particle's transverse momentum to estimate the probability of being a pileup~\cite{Catani:1993hr,Ellis:1993tq}.

Exploring the advantages of CHS and \textsc{SoftKiller}, we propose an optimal combination: a hybrid strategy named CHS+SK$_0$. This method first uses CHS to eliminate charged pileup particles, specifically targeting those with a vertex distance greater than $0.1\,\rm{mm}$. 
Following this, \textsc{SoftKiller} comes into play, 
removing pileup photons and neutral hadrons that fall below a certain transverse momentum threshold.
To avoid overcorrection, we have carefully configured \textsc{SoftKiller} to bypass charged hadrons.

Before showcasing the impressive performance of CHS+SK$_0$, it is necessary to outline the crucial simulation steps involved.
We need to make two important changes to the \textsc{Delphes} settings: 
first, we remove the pileup subtractors, and second, we turn off the \texttt{unique object finder} module. 
(However, we ensure that calculations for electron and muon isolation remain intact.)
Following these adjustments,  the refined output from \textsc{Delphes} is directed to a pileup subtraction module. 
In the final phase, jet clustering is executed.


\begin{figure}[!t]
\centering
\includegraphics[width=\textwidth]{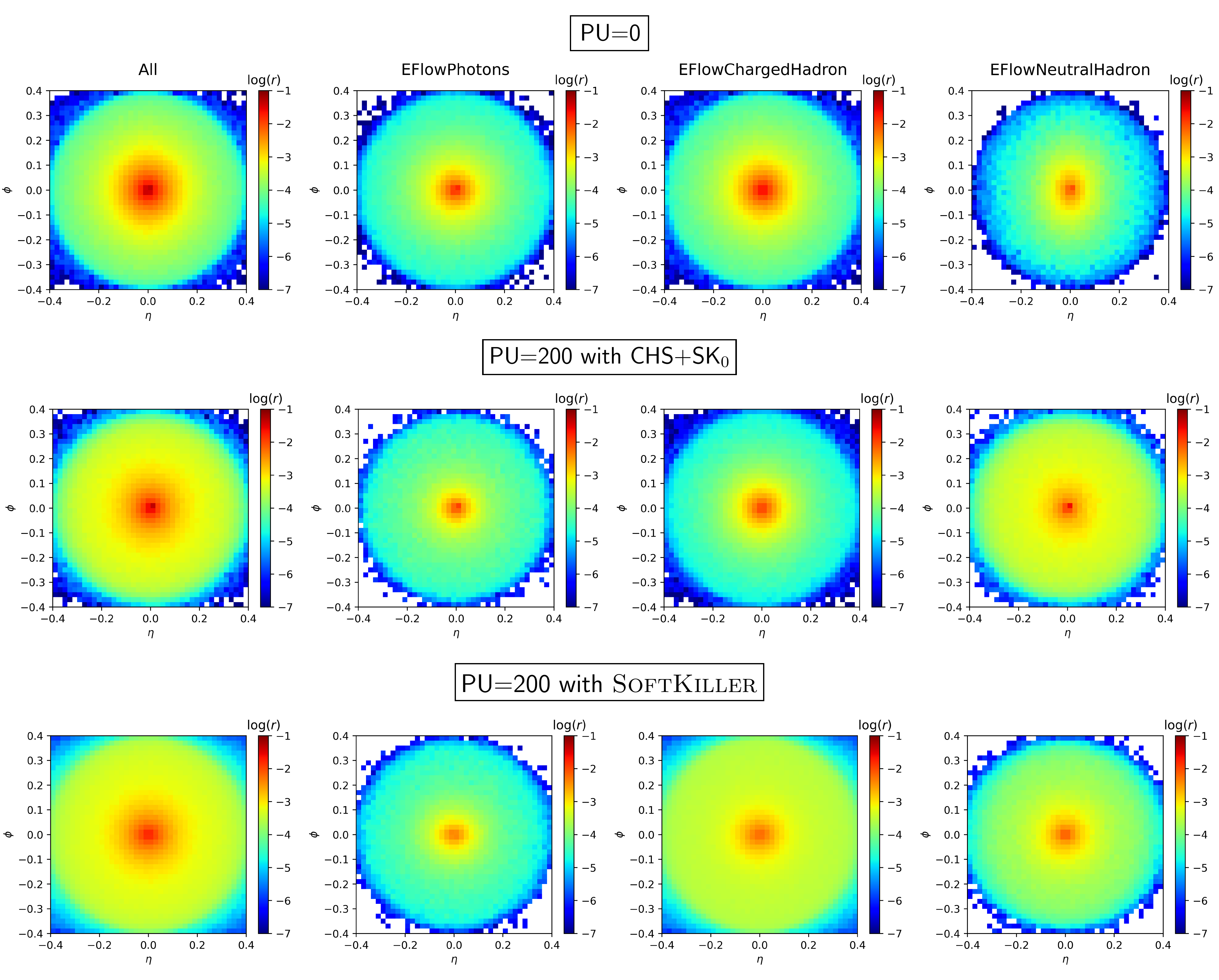}
\caption{
Jet images of the $\wpm jj$ background, 
where the color scale indicates the logarithm of the ratio of subparticle $\pt$ to the mother jet's $\pt$. 
We examine three pileup subtraction scenarios: no pileup (upper panels), 
200 pileups using the CHS+SK$_0$ subtraction method  (middle panels), 
and 200 pileups using the \textsc{SoftKiller}  (lower panels). 
The presentation spans four distinct jet image types: 
total jet images (first column), EflowPhotons (second column), EflowNeutralHadrons (third column), 
and EflowChargedHadrons (fourth column).
}
\label{fig-PU-jet-image-Wjj}
\end{figure}

We now turn our attention to demonstrating the exceptional performance of CHS+SK$_0$, utilizing jet images to provide a visual representation of the $\pt$ distribution of jet subparticles across a $\eta \times \phi$ grid. In \autoref{fig-PU-jet-image-Wjj}, jet images for the leading jet from the $\wpm jj$ background are presented, derived from an extensive sample of $10^5$ events.

Preprocessing involves translation and normalization techniques, as outlined in Ref.~\cite{Ren:2021prq}. 
We then sum the transverse momenta of all the subparticles within the jet and represent the intensity using $\log r$, 
where $r$ is the ratio of the subparticle's $\pt$ to the mother jet's $\pt$.
The $\log r$ information is distributed across the recalibrated $\eta$ and $\phi$ coordinates of each subparticle, which are now positioned relative to their mother jet. 
Here, we have adopted a pixel size of $\Delta\eta \times \Delta\phi = 0.02 \times 0.02$, reflecting the resolution of the simulated CMS electromagnetic calorimeter.

Equipped with these jet images, we are ready to conduct a comprehensive comparison between the CHS+SK$_0$ and \textsc{SoftKiller} subtraction methods.
In \autoref{fig-PU-jet-image-Wjj}, we explore three scenarios of pileups.
The top panels present jet images in the absence of pileups, providing a reference for our pileup subtraction efforts. 
The middle and bottom panels, on the other hand, display jet images with 200 pileups, processed using the CHS+SK$_0$ and \textsc{SoftKiller} subtraction methods, respectively.

Further breaking down our analysis, we categorize it into four distinct channels, each illustrated column-wise: 
total jet, EflowPhotons, EflowChargedHadrons, and EflowNeutralHadrons. 
As \autoref{fig-PU-jet-image-Wjj} vividly demonstrates, the CHS+SK$_0$ method significantly outperforms its counterpart, 
particularly in the efficient removal of charged pileup hadrons. 
This leads us to opt for the CHS+SK$_0$ subtraction technique for our subsequent analyses, 
incorporating all 200 pileup events.

Finally, we establish clear definitions for our terminology related to jets:
\begin{description}
\item[Jet ($J$):] 
A jet encompasses all physical entities that deposit energy in the calorimeters and undergo clustering by a jet algorithm. It is represented as $J$.
\item[Diphoton Jet ($\drr$):] 
A clustered jet is termed a diphoton jet if its two leading subparticles are EFlowPhotons. We denote this as $\drr$.
\item[QCD Jet ($j$):]
A QCD jet, stemming from quarks or gluons, is represented as $j$.
\item[Subparticle ($s_{ij}$):]
Each EFlow object inside a jet is referred to as a subparticle. The notation $s_{ij}$ denotes the $i$-th subparticle in the $j$-th jet. Both jets and subparticles are arranged in descending order of their $\pt$.
\end{description}

\section{Cut-Based Analysis}
\label{sec:cut-based}

\begin{table*}[!t]
\setlength\tabcolsep{10pt}
\centering
{\renewcommand{\arraystretch}{1.1} 
\begin{tabular}{|c||c|c|c|c|c|}
\toprule
BP no.   &  $\mhf$ & $\mach$ & $\sba$  & $m_{12}^2$ [GeV$^2$] & $\tb$   \\[3pt] \hline
BP-1        & \multirow{6}{*}{$1\gev$} & $150\gev$ & $-0.123$ & $0.0786$ & 8.06 \\ 
BP-2        & 			     & $175\gev$ & $-0.0909$ & $0.0400$ & 11.0 \\ 
BP-3        &  			     & $200\gev$ & $-0.0929$ & $0.0813$ & 10.7 \\ 
BP-4        &  			     & $250\gev$ & $-0.0941$ & $0.0494$ & 10.6 \\ 
BP-5        & 			      & $300\gev$ & $-0.0985$ & $0.0237$ & 10.1 \\ 
BP-6        & 			      & $331\gev$ & $-0.0974$ & $0.0634$ & 10.2 \\  \hline
BP-7        & \multirow{6}{*}{$5\gev$} & $150\gev$ & $-0.0737$ & $0.305$ & 13.5 \\ 
BP-8        & 			     & $175\gev$ & $-0.0922$ & $2.20$ & 10.8 \\ 
BP-9        &  			     & $200\gev$ & $-0.0983$ & $1.93$ & 10.1 \\ 
BP-10        &  			     & $250\gev$ & $-0.0907$ & $1.99$ & 11.0 \\ 
BP-11      & 			      & $300\gev$ & $-0.0984$ & $1.84$ & 10.1 \\ 
BP-12      & 			      & $331\gev$ & $-0.0920$ & $2.17$ & 10.8 \\  \hline
BP-13        & \multirow{6}{*}{$10\gev$} & $150\gev$ & $-0.0748$ & $1.17$ & 13.3 \\ 
BP-14        & 			     & $175\gev$ & $-0.0993$ & $1.70$ & 10.0 \\ 
BP-15        &  			     & $200\gev$ & $-0.0919$ & $0.973$ & 10.8 \\ 
BP-16        &  			     & $250\gev$ & $-0.0974$ & $0.851$ & 10.2 \\ 
BP-17      & 			      & $300\gev$ & $-0.0917$ & $0.0396$ & 10.9 \\ 
BP-18      & 			      & $328.3\gev$ & $-0.0979$ & $1.15$ & 10.2 \\  \hline
\end{tabular}
}
\caption{Benchmark points for the very light $\hf$.
All the parameter points satisfy the theoretical and experimental conditions.}\label{tab:BPs}
\end{table*}

In this section, we perform a signal-to-background analysis using the traditional cut-based approach. 
Our primary goal is to attain high signal significances across the entire parameter space for the very light $\hf$. 
To achieve this, we analyze 18 benchmark parameter points, as listed in Table \ref{tab:BPs}. 
For each signal benchmark point, we generate $3 \times 10^6$ events. 
Additionally, we consider the ten background processes specified in Table \ref{table:BG:xsec}. 
All events are processed through \textsc{Pythia8} and \textsc{Delphes}, 
employing the \textsc{Delphes} configuration outlined in the preceding section.

With our simulated data set ready, we implement the basic selection criteria as follows:
\bit
\item[$-$] There must be exactly one lepton with $p_T^\ell > 20\gev$ and $|\eta_\ell|<2.5$.
\item[$-$] The leading jet is required to satisfy $p_T^{J_1} > 50\gev$ and $|\eta_{J_1}|<2.5$.
\item[$-$] The subleading jet should fulfill the conditions $p_T^{J_2} > 30\gev$ and $|\eta_{J_2}|<2.5$.
\item[$-$] The missing transverse energy should exceed $\met>10\gev$.
\eit

In pursuit of optimizing signal significances, we highlight two distinguishing characteristics of our signal:
(i) the two leading subparticles in two leading jets are predominantly EFlowPhotons;
(ii) these leading subparticles contribute significantly to the transverse momentum of their mother jet.

\begin{figure}[!t]
    \centering
    \includegraphics[width=\linewidth]{fig-diphoton-tagging-P}
    \caption{$P(\hf \to \drr)$ and $P(j \to \drr)$ as functions of $p_T^J$,
    for the leading jet in the left panel and the subleading jet in the right panel.
    $P(\hf \to \drr)$ represents the probability of two photons from $\hf$ being identified as a diphoton jet, 
    while $P(j \to \drr)$ is the rate of a QCD jet tagged as a diphoton jet in the $\wpm jj$ background. 
 The red, green, and orange lines depict signal results for benchmark points BP-1, BP-7, and BP-13, respectively.
 }
    \label{fig-diphoton-tagging-P}
    \end{figure}

To highlight the first characteristic, 
we present the probabilities $P(\hf \to \drr)$ and $P(j\to \drr)$ against the $p_T$ of the mother jet 
in \autoref{fig-diphoton-tagging-P}.
Results for the leading and subleading jets are presented in the left and right panels, respectively.
$P(\hf \to \drr)$ represents the probability of the two photons from an $\hf$ decay being identified as a diphoton jet,
with the red, green, and orange lines corresponding to benchmark points BP-1, BP-7, and BP-13, respectively.
On the other hand, $P(j\to \drr)$ denotes the rate at which a QCD jet is misidentified as a diphoton jet in the $\wpm jj$ background.\footnote{A thorough analysis reveals that $P(j\to \drr)$ in the $Zjj$ background is similar to that in the $\wpm jj$ background,
within 10\%.}

For the signal, the probability $P(\hf\to\drr)$ remains substantial, consistently surpassing 40\% when $\pt^J \geq 50\gev$.
However, the relationship between this probability and $p_T^J$ varies with $\mhf$. 
For BP-7 ($\mhf=5\gev$) and BP-13 ($\mhf=10\gev$), 
the probability rises with increasing $p_T^J$, reaching approximately 85\%. 
In contrast, BP-1 ($\mhf=1\gev$) shows a distinct pattern: 
an initial increase, followed by a peak, and then a decrease as $\pt^J $ rises. 
This behavior can be attributed to the small $\mhf$ value in BP-1. 
Since  $R_\rr \sim 2 \mhf/\pt$, some of the two photons with high $\pt^J$ are so collimated 
that they nearly merge into a single EFlowPhoton, making them challenging to identify as a diphoton jet.
Nevertheless, the probability value even for BP-1 remains sizable,
hovering around 40\%.
On the other hand, the mistagging rate $P(j\to\drr)$ is only a few percent, demonstrating a clear distinction between signal and background.

The second salient feature of the signal is
the large ratios of $p_T$ of the two leading subparticles to the $\pt$ of their mother jet $J$.
In the case of the signal, the diphoton jet is mainly composed of two hard photons,
resulting in the leading and subleading subparticles holding a considerable share of $p_T^{J}$.
In contrast, a QCD jet consists of a diverse mix of particles, numbering from tens to well over a hundred. 
Consequently, it is rare for the two leading subparticles in a QCD jet to occupy a significant portion of $p_T^{J}$. 
To more vividly illustrate this distinction, we define:
\bea
\label{eq:ratio}
r_{ij} = \frac{p_T^{s_{ij}}}{p_T^{J_j}} \quad\text{for $i,j=1,2$}.
\eea

\begin{figure}[!t]
    \centering
    \includegraphics[width=\linewidth]{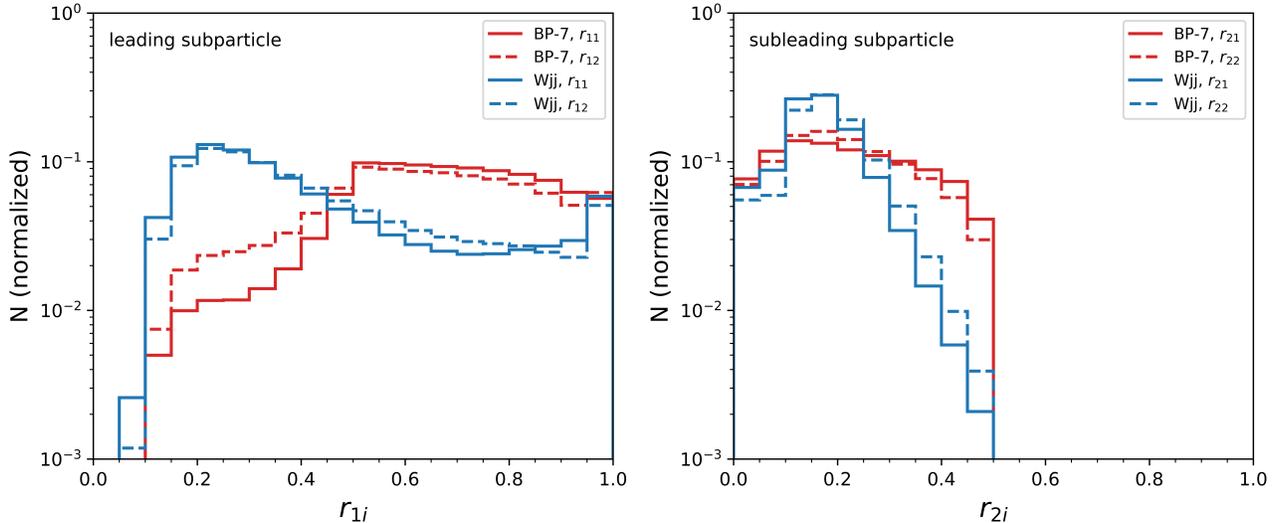}
    \caption{Normalized distributions of $r_{ij}$ for the signal in BP-7 (red) and the $\wpm jj$ background (blue)
    after the basic selection.
    Here, $r_{ij}$ is the $\pt$ ratio defined in \autoref{eq:ratio}. 
    The left panel presents the results for the leading subparticle, while the right panel focuses on the second-leading subparticle. Solid lines correspond to results for the leading jet, whereas dashed lines represent the subleading jet.
    }
    \label{fig-pt-ratio-dist}
    \end{figure}

To demonstrate this second feature, we present in \autoref{fig-pt-ratio-dist} the normalized distributions of $r_{ij}$ for both the signal BP-7 (in red) and the $\wpm jj$ background\footnote{Our 
analysis revealed that the $r_{ij}$ distributions in the $Zjj$ background closely resemble those of $\wpm jj$.}  (in blue) after the basic selection.
The left panel showcases the $\pt$ ratio for the leading subparticle, $r_{1i}$, 
while the right panel focuses on the subleading subparticle, $r_{2i}$. 
Solid lines depict results for $J_1$, and dashed lines correspond to $J_2$.

A primary observation reveals that the $r_{1i}$ value for the signal consistently surpasses 0.5, indicating that  the leading subparticle of a diphoton jet contributes almost half of its mother jet's $\pt$.
In contrast, the ratio for the $W^\pm jj$ background typically remains under 0.5.
Nevertheless, a noticeable peak around $r_{1i} \simeq 0.9$ in the $W^\pm jj$ background suggests
that merely imposing an upper bound on $r_{1i}$ may not sufficiently differentiate the signal from the background. 
Consequently, we shift our focus to the $r_{2i}$ distributions. 
While both the signal and background inherently exhibit $r_{2i}<0.5$, 
the signal's $r_{2i}$ is notably larger. 
By imposing a condition of $r_{2i}>0.25$, which corresponds to $r_{1i}<0.75$,  
we adeptly avoid the subtle peak around $r_{1i} \sim 0.9$ in the $W^\pm jj$ background.

\begin{table*}[!t]
\setlength\tabcolsep{10pt}
\centering
{\renewcommand{\arraystretch}{1.3} 
\begin{tabular}{|c||r||r|r|r|r||r|}
\toprule
\multicolumn{7}{|c|}{Cross sections in units of fb at the 14 TeV LHC  with $\lumtot=3\iab$}\\ \toprule
Cut                    &   \multicolumn{1}{c||}{BP-7} & \multicolumn{1}{c|}{$\wpm jj $}  & \multicolumn{1}{c|}{$Z jj$}    &  \multicolumn{1}{c|}{$\ttop$}	& \multicolumn{1}{c||}{$\wpm j \gm$}  	& \multicolumn{1}{c|}{$\mcs_{\rm BP-7}^{10\%}$} \\[3pt] \hline
Basic                & 34.8     & 372\,622     & 27\,727   & 32\,052 		& 3\,047 			& $1.09\times 10^{-3}$ \\[3pt] \hline
$\met>50\gev$ & 29.7     & 318\,407      & 23\,274  & 27\,395  		& 2\,610  			& $9.01\times 10^{-4}$ \\[3pt] \hline
$r_{11}>0.50$ & 24.9      & 102\,182      & 7\,843     & 4\,150 		& 1\,214 			&  $2.15\times 10^{-3}$ \\[3pt] \hline
$r_{12}>0.50$ & 18.7      & 36\,204        & 2\,853     & 692  		& 541 			& $4.56\times 10^{-3}$ \\[3pt] \hline
$r_{21}>0.25$ & 7.06      & 4\,218          & 323      & 62.2  			& 55.8 			& $1.49\times 10^{-2}$ \\[3pt] \hline
$r_{22}>0.25$ & 2.40      & 840            & 61.3     & 8.61  			& 10.1			& $2.56\times 10^{-2}$\\[3pt] \hline
$J_1 \to \drr$ & 2.29       & 18.6            & 2.31     & 0.205  			& 0.467			& 1.01 \\[3pt] \hline
$J_2 \to \drr$ & 1.98       & 0.363          & 0.0589 & 0.00 		  & 0.00849			& 22.8 \\[3pt] 
\bottomrule
\end{tabular}
}
\caption{Cross-section cut-flow chart for BP-7 and the main backgrounds from $\wpm jj$, $Zjj$, $\ttop$, and $\wpm j\gm$ at the 14 TeV LHC.
The presented cross sections are in femtobarns (fb). 
The basic selection criteria and the ratio $r_{ij}$ are detailed in the main text. 
For calculating the signal significance ($\mcs$), we take into account a 10\% background uncertainty and assume an integrated luminosity ($\lumtot$) of $3\iab$.
}\label{tab:cutflow}
\end{table*}

Based on the aforementioned two characteristics of the signal,
we devise a strategy to optimize the signal significance using a cut-based analysis. 
The cut-flow chart in Table \ref{tab:cutflow} outlines the cross sections for the signal and 
the four main backgrounds—$\wpm jj$, $Zjj$, $\ttop$, and $\wpm j\gm$—at the 14 TeV LHC. 
We have selected BP-7 as the representative benchmark point for detailed presentation, 
as it exemplifies the common trends observed across 18 benchmarks.
While we have comprehensively analyzed other backgrounds of Table \ref{table:BG:xsec}, 
they are omitted in Table \ref{tab:cutflow}  due to their negligible impact.

The final column in Table \ref{tab:cutflow} 
offers the signal significance $\mcs$, defined by~\cite{Cowan:2010js}:
\bea
\label{eq:significance:nbg}
\mathcal{S} = 
\left[2(\nsg + \nbg) \log\left(\frac{(\nsg + \nbg)(\nbg + \dbg^2)}{\nbg^2 + (\nsg + \nbg)\dbg^2} \right)  
- 
\frac{2 \nbg^2}{\delta_b^2} \log\left(1 + \frac{\dbg^2 \nsg}{\nbg (\nbg + \dbg^2)}\right)\right]^{1/2}.
\eea
Here, $\nsg$ denotes the number of signal events, 
$\nbg$ the number of background events, and $\dbg = \Dbg  \nbg$ the background uncertainty yield.
We take a 10\% background uncertainty ($\Dbg = 10\%$).

The results in Table \ref{tab:cutflow} are remarkable. 
After the basic selection, the four primary backgrounds overwhelm the signal, 
yielding the significance to an order of magnitude of $10^{-3}$. 
The cut on the missing transverse energy, pivotal for neutrino tagging, 
fails to boost the significance due to the presence of a neutrino in the dominant $\wpm jj$ background.
The differentiation becomes evident when applying the $\pt$ ratio cuts.
By enforcing $r_{11}>0.5$ and $r_{12}>0.5$, we retain approximately 63\% of the signal events that survive the $\met > 50\gev$ cut, while the backgrounds are diminished to $\mco(10^{-3})$.
Further imposing $\pt$ ratio conditions of $r_{2i}>0.25$ effectively suppresses the backgrounds.
Yet, the signal significance remains relatively low, hovering around 2.6\%. 

The last two selection criteria are decisive.
We first require that the leading jet must be a diphoton jet. 
While this condition significantly reduces the $\nsg/\nbg$ ratio, it is not enough to markedly elevate the significance. 
The final condition that the subleading jet also be a diphoton jet is what truly drives up the significance. 
When accounting for a 10\% background uncertainty, 
the final significance ascends to 22.8, affirming the discovery of a very light fermiophobic Higgs boson.

\begin{table*}[!t]
\centering
{\renewcommand{\arraystretch}{1.3} 
\begin{tabular}{|c|c|c||c|c|c||c|c|c|}
\toprule
\multicolumn{9}{|c|}{Results in the cut-based analysis at the 14 TeV LHC  with $\lumtot=3\iab$}\\ \toprule
 & ~~$\sg_{\rm final}$ [fb]~~ & ~~~$\mcs^{10\%}$~~~ &  & ~~$\sg_{\rm final}$ [fb]~~  & ~~~$\mcs^{10\%}$~~~ & & ~~$\sg_{\rm final}$ [fb]~~ & ~~~$\mcs^{10\%}$~~~ 
 \\[3pt]\hline
~~~BP-1~~~ 	& 1.46 	& 18.5 & ~~~BP-7~~~ 	& 1.98  	& 22.8 	& ~~~BP-13~~~ 	& 1.81 	& 21.5 \\[3pt]\hline
BP-2 		&  1.19 	& 16.1 & BP-8 			& 1.68 	& 20.4 	& BP-14 			& 1.56 	& 19.4 \\[3pt]\hline
BP-3 		& 0.927 	& 13.4 & BP-9 			& 1.37 	& 17.7 	& BP-15 			&1.29 	& 17.1  \\[3pt]\hline
BP-4 		& 0.529 	& 8.71 & BP-10 		& 0.900 	&  13.0 	& BP-16 			& 0.857 	& 12.7 \\[3pt]\hline
BP-5 		& 0.303 	& 5.49 & BP-11 		& 0.582 	& 9.40 	& BP-17 			& 0.566 	&  9.19  \\[3pt]\hline
BP-6 		& 0.216 	& 4.09 & BP-12 		& 0.457 	& 7.74 	& BP-18 			& 0.456 	& 7.72  \\[3pt]
\bottomrule
\end{tabular}
}
\caption{Signal cross sections and the significance values after the final selection at the 14 TeV LHC.
Calculations are based on a total integrated luminosity of $3\iab$ and a 10\% background uncertainty. 
}\label{tab:xsec:significance:cut}
\end{table*}

Moving forward, we present the conclusive results for all 18 benchmark points. 
Table \ref{tab:xsec:significance:cut} presents the signal cross sections  after the final selection
and the corresponding significance values at the 14 TeV LHC. These computations are based on a total integrated luminosity of $3\iab$ and a 10\% background uncertainty.  
The comprehensive suite of cuts in Table \ref{tab:cutflow} is uniformly applied across all benchmark points, 
avoiding tailored adjustments for specific benchmark points in pursuit of unbiased analysis.
The significance values obtained are encouraging. 
With the exception of BP-6, every benchmark point boasts significance values surpassing 5. 
Even the notably challenging BP-6 achieves a respectable significance of 4.09.

We observe distinct trends in significance depending on the benchmark points.
When holding $\mhf$ constant, the significance tends to decrease as $\mch$ increases, 
a reduction primarily due to the smaller signal cross section from the limited kinematic phase space available at higher $\mch$ values. 
Conversely, when fixing $\mch$, scenarios with $\mhf=5\gev$ consistently yield the highest significances. 
The slightly reduced significances observed in scenarios with $\mhf=10\gev$
result from a subset of signal events producing two photons with $\Dt R>0.4$, 
which fails to satisfy the criteria for two diphoton jets.
On the other hand, scenarios featuring $\mhf=1\gev$ consistently exhibit the lowest significance values. 
This small $\mhf$ leads to two collimated photons, 
causing a significant portion of signal events to not satisfy the two diphoton-jet requirement.

\section{Mass Reconstruction for $\mhf$ and $\mch$}
\label{sec:mass:reconstruction}

In the previous two sections, 
we underscored the efficacy of our cut-based analysis strategy in achieving robust significance values. 
Our next aim is to validate that the observed signal indeed originates from the $pp\to \hf\ch\to \hf\hf\wpm$ process. 
Precisely identifying the source necessitates the reconstruction of $\mhf$ and $\mch$.
Since $\hf$ predominantly decays into two photons, 
$\mhf$ can be reconstructed using the invariant mass of the two photons within a diphoton jet. 
To reconstruct the mass of the charged Higgs boson,  we focus on $\mtch$, the transverse mass of $\ch$ as it decays to $\rr\ell\nu$. 

To initiate the calculation of $\mtch$, we first define the four-momentum of the visible components, denoted as $p_{\rm vis}^\mu$. 
A challenge arises due to the presence of an additional diphoton jet in the full scattering process, 
leading to ambiguity in determining which diphoton jet results from the $\ch$ decay. 
To navigate this, we adopt a reasonable assumption: the diphoton jet stemming from the decay of $\ch$ 
is the subleading jet. 
This assumption is based on the observation that the prompt diphoton jet generally exhibits a higher $\pt$ 
than the one involved in the decay chain.\footnote{Our preliminary simulations indicate an approximate 20\% contamination resulting from this assumption.}

Following this assumption, we establish:
\bea
\label{eq:pvis}
p_{\rm vis}^\mu = p_{s_{12}}^\mu + p_{s_{22}}^\mu  +p_{\ell}^\mu. 
\eea
The square of the transverse mass of the charged Higgs boson is then defined as:
\bea
\label{eq:mtch}
\lf \mtch \ri^2= m_{\rm vis}^2
+ 2\left[
E_T^{\rm vis} \met - \vec{p}_T^{\rm \; vis} \cdot\vec{E}_T^{\rm \, miss}
\right]
,
\eea
where $m_{\rm vis}^2 = p_{\rm vis} \cdot p_{\rm vis}$, $E_T^{\rm vis} = \sqrt{m_{\rm vis}^2 + \lf {p}_T^{\rm vis} \ri^2}$,
and $\vec{E}_T^{\rm \, miss} = - \sum_i \vec{p}_T^{\, i}$ with $i$ covering all the observed particles
after the pileup subtraction.

In our endeavor to determine $\mhf$ and $\mtch$, we are confronted with a formidable challenge: 
obtaining accurate \emph{background} distributions \emph{after imposing the final selection}.
While the final selection guarantees robust signal significances through a drastic reduction in background events---leaving only 51 events for the $\wpm jj$ background and 4 events for $Zjj$---this scarcity of events  impairs our ability to acquire precise distributions for both $\mrr$ and $\mtch$. 
However, abandoning the final selection is not an option,
as the second-to-last cut results in an unacceptably low significance, falling below one. 
Furthermore, intensifying event generation to amplify the number of background events is impractical, 
as our computational resources are already maximized, 
with $5\times 10^8$ events generated for $\wpm jj$ and $5\times 10^7$ for $Zjj$.

In tackling this challenge, we have developed a novel approach that incorporates the mistagging probability $P(j \to \drr)$ as a weighting factor, a method we term the Weighting Factor Method (WFM). 
To grasp the benefits of WFM, it is instructive to examine the methodology of traditional cut-based analyses.
These analyses operate by either retaining or discarding events based on selection criteria, 
effectively assigning a binary weight of one or zero to each event. 
While straightforward, this method proves inefficient for analyzing background distributions
when the selection efficiency is exceedingly low. 
For instance, the final selection efficiency for the $\wpm jj$ backgrounds, relative to the basic selection, is an astonishingly sparse $10^{-7}$.

In contrast, our WFM strategically utilizes the continuous nature of the weighting factor $P(j \to \drr)$.
This approach ensures the inclusion of nearly all pertinent background events, 
ensuring a thorough representation of the background. 
For a comprehensive explanation of WFM, including a detailed discussion on how we model $P(j \to \drr)$, please refer to Appendix \ref{appendix:WFM}.

\begin{figure}[!t]
    \centering
    \includegraphics[width=0.95\linewidth]{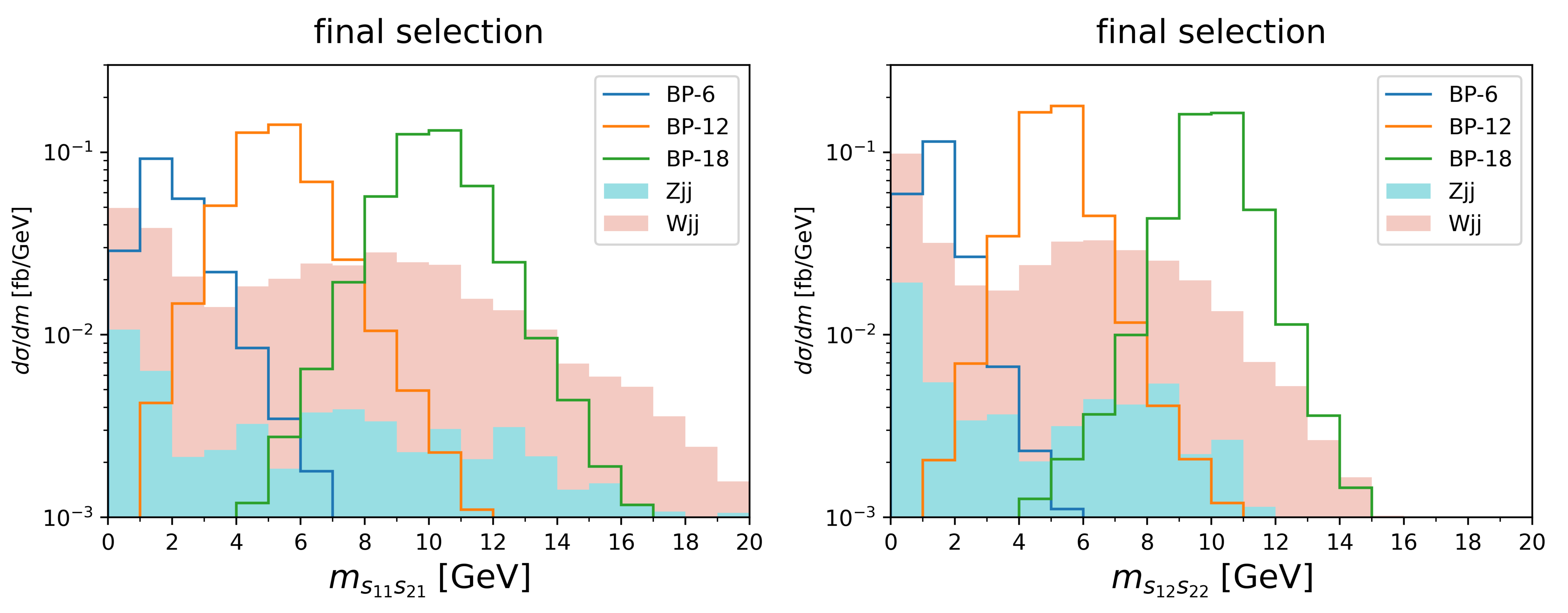}
    \caption{Invariant mass distributions for the two leading subparticles in the leading jet (left panel) and the subleading jet (right panel) at the 14 TeV LHC. All depicted events meet the final selection criteria.
    For the stacked $\wpm jj$ and $Z jj$ backgrounds, the WFM is utilized. 
    The expected signals for BP-6 (blue), BP-12 (orange), and BP-18 (green) are illustrated with solid lines.
 }
    \label{fig-mrr}
    \end{figure}

In \autoref{fig-mrr}, we depict the distributions of $m_{s_{11}s_{21}}$ (left panel) and $m_{s_{12}s_{22}}$ (right panel) 
for both signal and background events at the 14 TeV LHC, 
adhering to the final selection criteria detailed in Table \ref{tab:cutflow}. 
We consider three signal benchmark points with a heavy $\mch$: BP-6 (blue), BP-12 (orange), and BP-18 (green). 
For the signal distributions, we rely on the results from the traditional cut-based analysis, justified by the ample number of signal events remaining post final selection. 
In addition, we display the results for the two primary backgrounds, $\wpm jj$ and $Z jj$,
in the stacked format,
using the WFM. 
We omit other backgrounds here due to their inconsequential contributions following the final selection.

A salient characteristic for the signal in \autoref{fig-mrr} is a distinct resonance peak,
for both the leading and subleading jets. 
This peak closely corresponds to the mass of the fermiophobic Higgs boson.
Conversely, the background distributions exhibit two peaks: a sharp one and a more diffuse secondary one. 
The acute peak, centered at $\mrr \simeq 0$, is predominantly attributed to light mesons, such as $\pi^0$, $\rho$, $\eta$, and $\eta'$, which decay into two photons.\footnote{We confirm this interpretation through our thorough analysis, 
which demonstrates the absence of a sharp peak without the diphoton jet criteria.}
Meanwhile, the broader peak around $\mrr\simeq 10\gev$ emerges 
as background events increasingly mimic the signal after meeting all selection criteria.
Nevertheless, 
the resonance peaks in the diphoton invariant mass distributions are clearly distinguishable from the backgrounds. 

\begin{figure}[!t]
    \centering
    \includegraphics[width=0.6\linewidth]{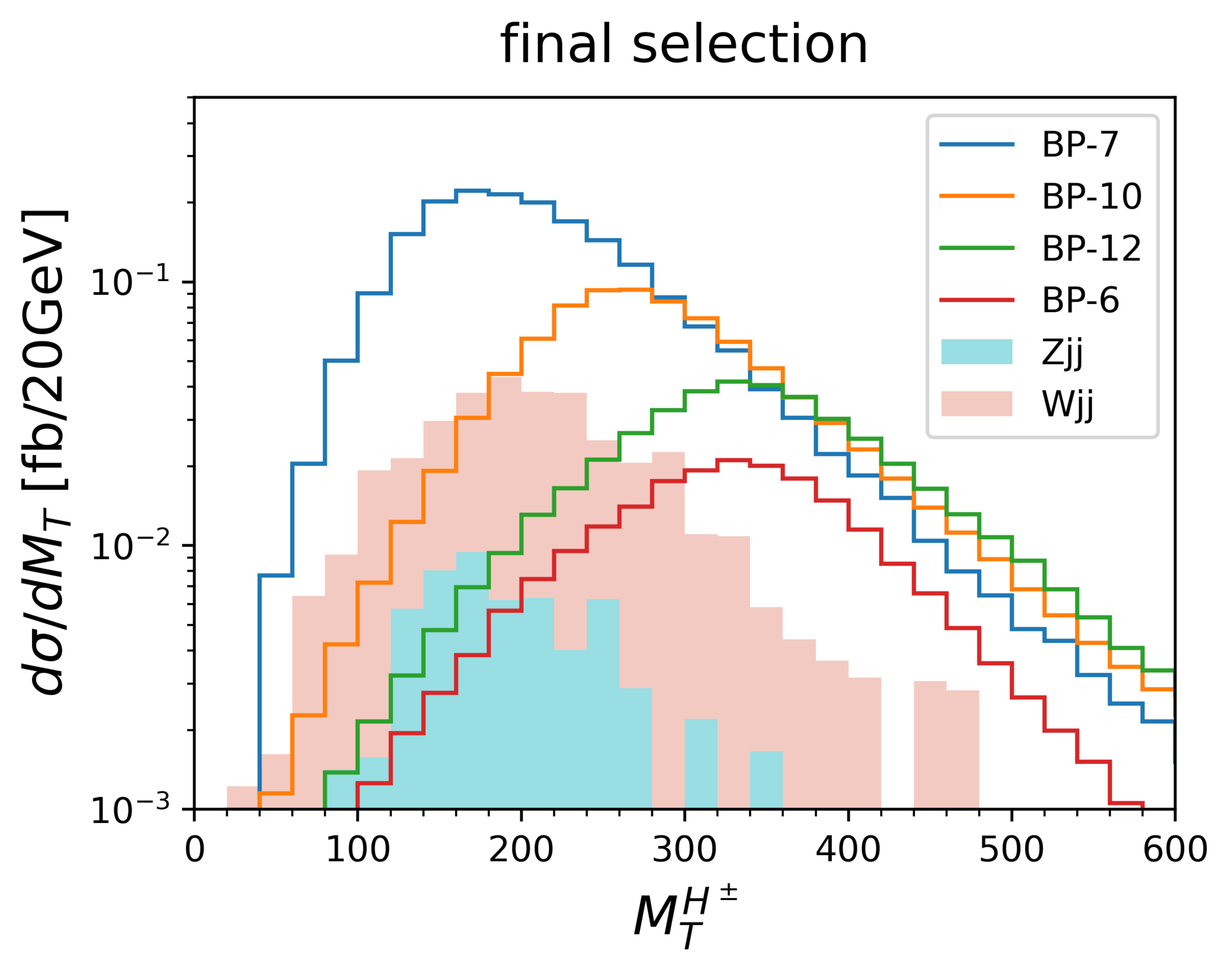}
    \caption{Distributions of the transverse mass of the charged Higgs boson after the final selection at the 14 TeV LHC. 
    The results for the $\wpm jj$ and $Z jj$ backgrounds are displayed in a stacked manner.
    The expected signals for BP-6 (red), BP-7 (blue), BP-10 (orange), and BP-12 (green) are represented by solid lines.
 }
    \label{fig-mT}
    \end{figure}
    
In \autoref{fig-mT}, we show the transverse mass distribution of the charged Higgs boson. 
For the signal, four benchmark points are considered: BP-6 (red), BP-7 (blue), BP-10 (orange), and BP-12 (green).
Additionally, we showcase the WFM results for the two primary backgrounds, $\wpm jj$ and $Zjj$, in a stacked manner.

The $\mtch$ distributions for the signal exhibit a unique wedge-shaped peak, 
marked by a sudden drop around $\mtch \simeq \mch$. 
However, this peak is broader than the well-known distribution shape of $\wpm\to \ell^\pm \nu$.
This broadening arises from two main factors. 
First, the long decay chain of $\ch\to \hf\wpm \to \rr \ell \nu$ introduces inherent uncertainties, 
especially when measuring the three momenta of the two photons and one lepton. 
Second, there is ambiguity in determining which diphoton jet originates from the $\ch$ decay. 
Despite the broadness, the characteristic shape of the transverse mass distribution is evident in the signal.

On the other hand,  the backgrounds show a single, broad hill-shaped peak centered around $180\gev$. 
This shape evolves as background events increasingly resemble the signal after satisfying all selection criteria. 
One might worry that the background peak around $180\gev$ could obscure the signal $\mtch$ peak 
when $\mch$ is close to 180 GeV. 
However, as indicated in \autoref{tab:xsec:significance:cut}, 
the significance values for $\mch=175\gev$ are so high that the $\mtch$ signal peaks remain distinct and easily distinguishable from the background contributions.

In conclusion, the mass reconstruction of $\mhf$ and $\mch$ 
is feasible,
signifying that the combined $\mrr$ and $\mtch$ distributions effectively and distinctly pinpoint the origin of our new signal.
 
\section{Machine Learning Approach for heavy $\mch$}
\label{sec:CNN}

In the previous two sections, 
we underscored the efficacy of our cut-based analysis strategy in achieving robust significance values as well as 
the mass reconstruction of $\mrr$ and $\mch$.  
Yet, challenges manifested when addressing the heavy charged Higgs boson.
For instance, BP-6 reached a significance of 4.09, 
which is not convincing enough to confirm the presence of the very light fermiophobic Higgs boson.    
Hence, in this section, we employ machine learning techniques, with a keen focus on BP-6, BP-12, and BP-18, 
aiming to enhance the significances.
At the parton-level, the total cross sections for these benchmarks are
$\sg_\tot (\text{BP-6})=9.62\fb$,
$\sg_\tot (\text{BP-12})=9.63\fb$, and
$\sg_\tot (\text{BP-18})=9.83\fb$.

Let us begin by discussing the preparation of input features.
We formulate two distinct features: the event feature and the subparticle feature.
The event feature comprises 21 elements, constructed as follows:
\begin{align}
\evfeat  = & 
\left[
p_T^{J_1}, \eta_{J_1}, \phi_{J_1}, m_{J_1}, p_T^{J_2}, \eta_{J_2}, \phi_{J_2}, m_{J_2}, p_T^{\ell}, \eta_{\ell}, \phi_{\ell}, \met,
\phi_\vmet,
\right.
\\ 
& \left.
\Dt R_{J_1J_2},\Dt R_{J_1\ell},\Dt R_{J_2\ell},\Dt R_{J_1\vmet},\Dt R_{J_2\vmet},\Dt R_{\ell\vmet},M_{T}^{J_1},M_T^{J_2} 
\right],
\end{align}
with $M_{T}^{J_i}$ ($i=1,2$) representing the transverse mass in \autoref{eq:mtch} 
using $p_{\rm vis}^\mu = p_{J_i}^\mu +p_{\ell}^\mu$.
For normalization, the feature elements with a mass dimension are divided by $500\gev$.
This list includes the transverse momentum $p_\text{T}$, the invariant mass $m_{J_i}$,
the missing transverse energy $\met$, and the transverse mass $M_{T}^{J_i}$.

The subparticle feature is divided into two vectors associated with $J_1$ and $J_2$. 
Each $J_i$ category includes the 10 leading subparticles, each characterized by three attributes: $p_T$, $\eta$, and $\phi$. 
As a result, the total dimension of the subparticle feature is $30 \times 2$.
The coordinates $\eta$ and $\phi$ of a given subparticle are adjusted to be relative to their mother jet.
We divide the $p_T$ values by $100\gev$ for normalization. 
To emphasize the photons, other particles (hadrons) are assigned a value of zero for all three attributes.

\begin{figure}[!t]
    \centering
    \includegraphics[width=0.86\linewidth]{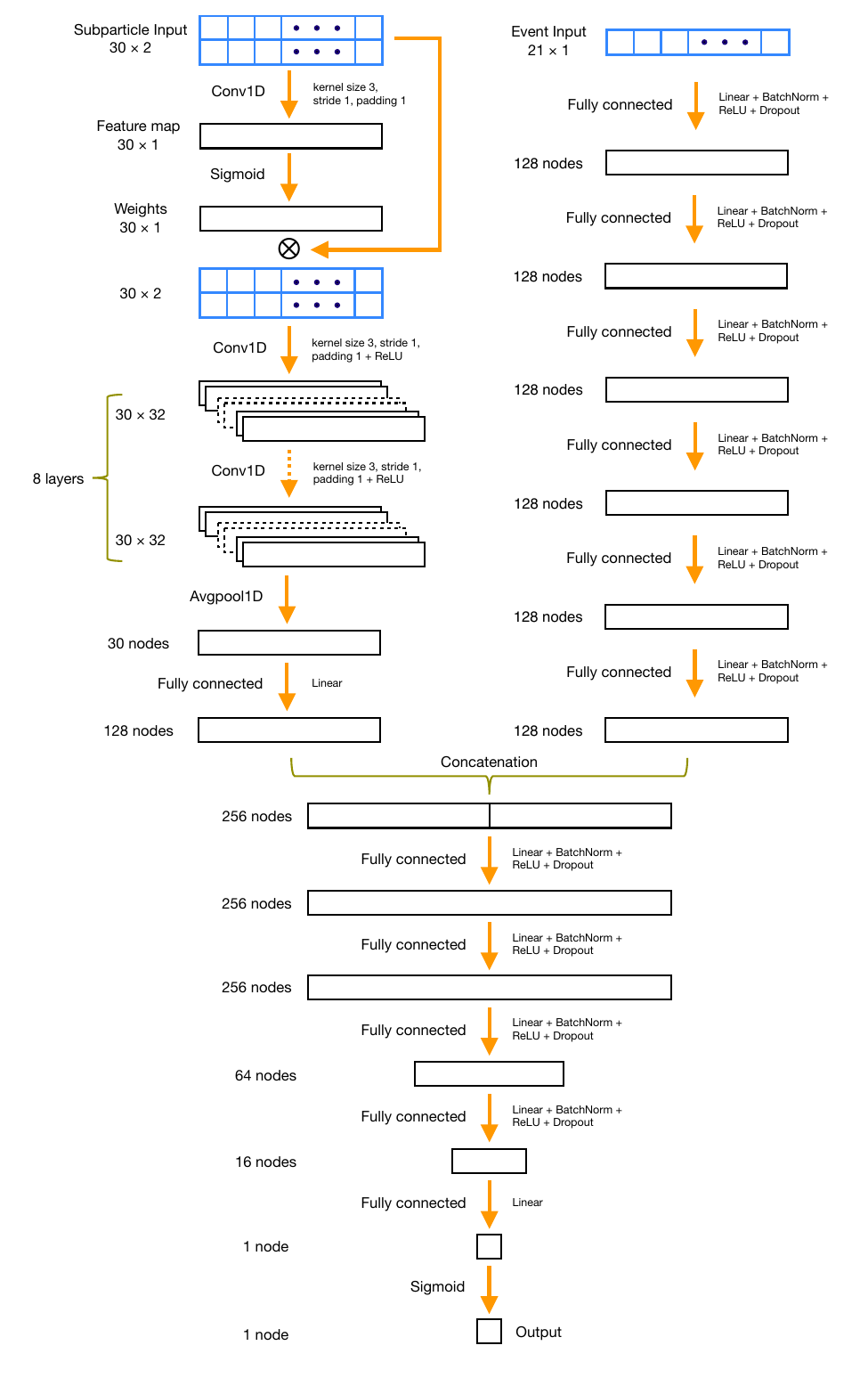}
    \caption{Model architecture of 1D CNN}
    \label{fig-CNN}
    \end{figure}

Our network architecture, illustrated in \autoref{fig-CNN}, 
consists of three main components:  a one-dimensional (1D) CNN block and two multilayer perceptrons (MLP1 and MLP2). 
The 1D CNN block is responsible for processing the subparticle feature, whereas MLP1 handles the event feature.  
MLP2 merges the outputs from both the 1D CNN and MLP1  to produce the final model prediction.
For those interested in the datasets and the detailed operation of the deep learning model, 
we have made them available on our GitHub repository.\footnote{\texttt{https://github.com/chofchof/light-hf-ml/}}

Diving into the details, the 1D CNN block comprises nine 1D convolutional layers. 
The first layer uses a kernel size of 3 and its output goes through a sigmoid function, 
which maps the values between 0 and 1. 
Functioning as attention weights,  these values are multiplied by each subparticle input feature.
This is a crucial step with a clear purpose: 
 it assigns varying weights to each element within the input features, thereby enhancing the model’s ability to focus on informative parts of the data.
The next eight layers are also 1D convolutions with a kernel size of 3, but they include a ReLU activation function to add non-linearity to the model. Following these layers, an average pooling operation and a fully connected layer condense the information into a 128-dimensional feature vector.

MLP1 primarily transforms the event input feature into a 128-dimensional feature vector. 
This perceptron comprises six fully connected layers, each containing 128 nodes.
Following each layer, batch normalization, a ReLU activation function, and a dropout layer with a 50\% probability are applied.

MLP2 finally determines the probability that an event is classified as a signal. 
Its architecture includes five fully connected layers with node counts of 256, 256, 256, 64, and 16, in succession. 
Each layer is followed by batch normalization, a ReLU activation function, and a dropout layer with a 50\% probability. 
After these five layers, an additional fully connected layer is set in place to produce an one-dimensional feature vector.  
This vector then undergoes processing via a sigmoid function, yielding the final classification probability as the output.

For optimal model implementation and precision, 
we utilize the renowned PyTorch deep learning framework~\cite{paszke2019pytorch}. 
Both training and evaluation processes are expedited using the NVIDIA Titan V GPU. 
We optimize model parameters with the AdamW optimizer~\cite{loshchilov2017decoupled}, 
which is set with an initial learning rate of 0.002 and a weight decay of 0.01, based on mini-batches of 512 training samples. Throughout the training phase, which spans 100 epochs, 
we decrease the learning rate by half every 10 epochs to enhance convergence. 

Now let us describe the generation and assignment of our dataset for training and evaluation. 
To leverage the unique attributes that differentiate the signal from the backgrounds, 
we enforce additional conditions $p_T^{J_1}>100\gev$ and $p_T^{J_2}>80\gev$,
after the basic selection. 
During the training phase, 
we employ training and validation datasets, each brimming with $6 \times 10^5$ events.
These datasets are evenly split for signals and backgrounds.  
The signal events are equally divided amongst BP-6, BP-12, and BP-18.
For the background events, which originate from ten processes, allocation is proportionate to their respective cross sections.

Central to our training and evaluation processes is the design of our loss function.
Our primary goal of enhancing detection significance 
necessitates efficient background rejection. 
Accordingly,  we have tailored the loss function to inversely correlate with signal significance.
For the sake of computational efficiency, we employ $1/Z$ as the loss function,
where  $Z$ is a concise representation for the significance:
\begin{equation}
\label{eq:significance:Z}
    Z = \frac{\nsg}{\sqrt{\nbg+\dbg^2 }},
\end{equation}
where we take into account a 10\% background uncertainty, denoted as $\dbg=0.1 \nbg$.

Upon concluding the training process, we extract the model’s optimal parameters and apply them to our entire test dataset---totaling $1.27 \times 10^8$ events, consisting of $9 \times 10^6$ signal events and an overwhelming $1.18\times 10^8$ background events. Subsequently, we apply a specified selection threshold $\xc$ on the outputs of all the test samples. Finally we then determine the comprehensive significance metric $\mcs$ in \autoref{eq:significance:nbg}.

Given two threshold options, $\xc=0.5$ and $\xc=0.9$, we present the signal significances for BP-6, BP-12, and BP-18 as follows:\beq
\label{eq:final:significances:CNN}
\begin{split}
\xc=0.5
&:\qquad
\mcs_{\text{BP-6}}^{10\%}  = \phantom{1}9.0,
\quad
\mcs_{\text{BP-12}}^{10\%}  = 15.4 , 
\quad
\mcs_{\text{BP-18}}^{10\%}  =15.0;
\\
\xc=0.9
&:\qquad
\mcs_{\text{BP-6}}^{10\%}  =  18.9,
\quad
\mcs_{\text{BP-12}}^{10\%}  = 33.2, 
\quad
\mcs_{\text{BP-18}}^{10\%}  =32.4.
\end{split}
\eeq
The outcomes from our CNN machine learning approach are indeed outstanding. 
Even with the conservative threshold of $\xc=0.5$, BP-6
now reaches a significance of $9.0$. 
Furthermore, both BP-12 and BP-18 witness approximately 100\% increases in their significances 
when compared to the results from the cut-based analysis. 
Opting for the more aggressive threshold of $\xc=0.9$ yields even more enhanced significances.
Collectively, these outcomes emphatically demonstrate the effectiveness of our model architecture.

\section{Conclusions}
\label{sec:conclusions}

We have comprehensively studied the phenomenological signatures 
associated with a very light fermiophobic Higgs boson $\hf$ 
with a mass range of $\mhf\in[1,10]\gev$ at the 14 TeV LHC. 
The light $\hf$ is postulated under the condition $\al=\pi/2$ within the inverted Higgs scenario of the type-I two-Higgs-doublet model.
Through an exhaustive scan of the parameter space, 
taking into account theoretical requirements, experimental constraints, and the cutoff scale  exceeding 10 TeV, 
we demonstrated that the $\mhf\in[1,10]\gev$ range retains  a substantial number of viable parameter points. 
This is largely attributed to the experimental complexities of detecting the soft decay products of $\hf$. 
Importantly, this mass range results in strictly defined parameter space, 
ensuring predictable phenomenological signatures. 
Two standout features of the viable parameter space are:
(i) the BSM Higgs bosons have a single dominant decay mode, such as $\hf\to\rr$, $\ch\to\hf\wpm$, and $A \to \hf Z$;
(ii) $\mch$ and $\ma$ are relatively light below $ \lsim 330\gev$.
Building on these insights, we have proposed a \emph{golden channel}, $pp\to\hf\ch\to \rr\rr \ell\nu$, for exploration of $\hf$ at the HL-LHC.

A serious challenge surfaces as the two photons from $\hf\to\rr$ fail to meet the photon isolation criteria, 
due to their high collimation within $\Dt R<0.4$. 
As a result, the final state (characterized by four photons) usually manifests as two jets, thereby facing immense QCD backgrounds.
To address this, we shifted our focus to the subparticles within the jet, 
identifiable as EFlow objects within the \textsc{Delphes} framework. 
This approach facilitates the extraction of information about a subparticle's type 
(EflowPhoton, EflowNeutralHadrons, or EflowChargedHadrons), 
subsequently enabling the probing of diphoton jets. 
The challenges posed by pronounced pileups, which could blur the distinction between diphoton jets and QCD jets,
are effectively addressed by our innovative pileup subtraction 
method---a hybrid solution combining charged hadron subtraction with \textsc{SoftKiller}.

With the method of probing diphoton jets,
we performed the full simulation for signal-to-background analysis at the detector level across 18 benchmark points.  
A universal strategy
was articulated for the cut-based analysis, yielding encouraging outcomes. 
Except for BP-6, characterized by $\mhf=1\gev$ and $\mch=330\gev$, 
all benchmark points exhibited signal significance considerably above 5. 
For the mass reconstructions of the BSM Higgs bosons, 
we analyzed both the invariant mass distribution of the two leading subparticles 
and the transverse mass of the charged Higgs boson, 
based on events post the final selection. 
Distinct peaks correlating with $\mhf$ and $\mch$ were prominently discerned above the background signals.
An inherent challenge---securing reliable background distributions with the scarce events post the final selection---is addressed through our pioneering Weighting Factor Method (WFM).

To cover the more challenging regions marked by a heavy charged Higgs boson mass, 
we employed machine learning techniques. 
A potent network structure was designed,
comprised of  a one-dimensional (1D) CNN block followed by two multilayer perceptrons. 
The efficacy of this model was commendable. 
With the nominal threshold of $\xc=0.5$, we managed to nearly double the significances for the heavy $\mch$ cases.

In this extensive research, we have explored uncharted territories 
 of a very light fermiophobic Higgs boson via diphoton jets. 
Our approach, harmonizing traditional analyses with innovative methodologies like hybrid pileup subtraction, 
the WFM, and machine learning, offers novel contributions to  the field. 
We urge the community to consider our findings in the quest for BSM signals.
 
\section*{Acknowledgments}
The work of J.C. is supported by National Institute for Mathematical Sciences (NIMS) grant funded by the Korea government (MSIT) (No.~B23810000). 
And the work of D.W., J.K., P.S., and J.S. is supported by
the National Research Foundation of Korea, Grant No.~NRF-2022R1A2C1007583.
The work of S.L. is supported by Basic Science Research Program through the National Research Foundation of Korea(NRF) funded by the Ministry of Education(RS-2023-00274098).

\appendix

\section{Weighting factor method}
\label{appendix:WFM}

In this appendix, we elaborate on the Weighting Factor Method (WFM).
Our focus sharpens on the modeling of $P(j\to \drr)$ for background processes, 
where $P(j\to \drr)$ represents the probability of a QCD jet misidentified as a diphoton jet.
The extreme scarcity of background events that pass the final selection criteria makes this approach crucial 
for attaining reliable distributions of $\mrr$ and $\mtch$,
which necessitate a substantial number of events.
Our discussion in this Appendix focuses on the dominant $\wpm jj$ backgrounds, 
considering that the next dominant $Zjj$ backgrounds contribute to only about 10\% of the $\wpm jj$ events.\footnote{Our 
rigorous analysis affirmed that the performance of the WFM for the $Zjj$ backgrounds is similar to that for 
$\wpm jj$.}

For clarity in our subsequent discussions, we elucidate some terminologies.
The expected number of events corresponding to a specific luminosity is denoted by $N$. 
In realistic simulations, however, the actual number of generated background events is less than $N$. 
For distinction, we denote it by $n$.
To be more explicit, let us define $E_\cut$ as the set of events that fulfill a certain \enquote{cut}. 
The number of events meeting this cut is determined by the cardinality of the set $E_\cut$:
\bea
\label{eq:ncut:def}
n_\cut \equiv \# E_\cut.
\eea
In the conventional cut-based analysis, the cross section after the final selection is then given by
\bea
\label{eq:xsec:cut-based}
\sg_\final^\text{cut-based} = \sum_{e \in E_\final}  1 \times \frac{ \sg_\tot}{n_{\rm gen}}  = \frac{n_\final}{n_{\rm gen}} \; \sg_\tot,
\eea
where $\sg_\tot$ represents the total cross section at the parton level.

Let us revisit the cut-flow presented in Table \ref{tab:cutflow}.
Following the basic selection, we have an accumulative sequence of criteria:
(i) $\met>50\gev$;
(ii) $r_{11}>0.5$;
(iii) $r_{12}>0.5$;
(iv) $r_{21}>0.25$;
(v) $r_{22}>0.25$;
(vi) $J_1 \to \drr$;
(vii) $J_2 \to \drr$.
The $\wpm jj$ backgrounds register counts of $n_{r_{22}} = 1.180 \times 10^5$ and $n_{\final} = 51$,
where the condition ${r_{22}} $ represents the accumulated conditions leading up to $r_{22}>0.25$.

Now we unpack how the WFM modifies $\sg_\final^\text{cut-based}$.
Instead of focusing on the background events post the final selection,
we shift our attention to the more extensive dataset refined by the $r_{22}$ condition.
For each background event $e$ within the set $E_{r_{22}}$, 
we determine $P_e(j_1\to \drr)$ and $P_e(j_2\to \drr)$ that serve as weight factors.
To compute the joint probability using these multipliers, we adopt an assumption:
the observation of $j_1$ as a diphoton jet remains statistically decoupled from $j_2$'s categorization.   
This implies that scenarios in which both jets are tagged as diphoton jets are derived 
from the multiplication of their respective weighting factors. 
Therefore, the cross section following the final selection, under the WFM framework, becomes
\bea
\label{eq:xsec:wtm}
\sg_\final^\text{WFM} = \sum_{e \in E_{r_{22}} }  P_e(j_1\to \drr) P_e(j_2\to \drr)  \times  \frac{ \sg_\tot}{n_{\rm gen}}.
\eea
It is important to reiterate: for $\sg_\final^\text{cut-based}$ in \autoref{eq:xsec:cut-based}, we consider the $n_\final$ events,
while for $\sg_\final^\text{WFM}$ in \autoref{eq:xsec:wtm}, we employ the $n_{r_{22}}$ events.

To model $P_e(j\to \drr)$ in practice, 
we need to compute the ratio of event counts after the $j\to \drr$ cut to those satisfying the $r_{22}$ cut. 
Recognizing that $P_e(j\to \drr)$ would naturally depend on event-specific characteristics like $p_T^j$, 
it is pertinent to focus on the event counts within a defined kinematic bin when calculating the ratio. 
Considering that the magnitude of $P_e(j\to \drr)$ is on the order of a few percent, a substantial volume of events that satisfy the $r_{22}$ condition must be collected in the reference set. 
Strategically, we adopt two-dimensional kinematic bins.\footnote{The binning strategy for $m_\rr$ and $\mtch$ distributions varies. For the invariant mass distribution of the two foremost subparticles in the QCD jet $j_{1,2}$, we employ the scheme $ (m_{s_{1i} s_{2i}}, p_T^{j_i})$, with $i$ taking values 1 or 2. In contrast, the $\mtch$ distribution utilizes the pair $ (\mtch, p_T^{j_i})$.}

For any specific event $e$, we introduce $B_e$ as the set of all events within the bin containing $e$.
Consequently, the probability of a QCD jet being incorrectly identified as a diphoton jet in event $e$ is given by:
\bea
\label{eq:P:mistagging:def}
P_{e} (j\to \drr) = 
\frac{\# \lf E_{j\to\drr} \cap B_e\ri }
{\# \lf E_{r_{22}} \cap B_e\ri },
\eea
where  the criteria within $E_{j\to\drr}$ means the combination of the $j\to\drr$ condition
with the $r_{22}$ cut.

The advantages of WFM become clear when analyzing kinematic distributions. 
As an illustration, consider a case where $\# \lf E_{r_{22}} \cap B_e \ri=1000$. 
Using the traditional cut-based method and implementing both $j_1 \to \drr$ and $j_2 \to \drr$ conditions, 
most of the kinematic bins become empty since the joint probability is exceedingly low as  $3.8\times 10^{-4}$. 
It is not feasible to obtain a reliable kinematic distribution in this case. 
In contrast, utilizing the WFM method, we can expect a projection of roughly 0.38 events post-final selection, 
enabling reliable distributions.

\begin{figure}[!t]
\centering
\includegraphics[width=0.65\textwidth]{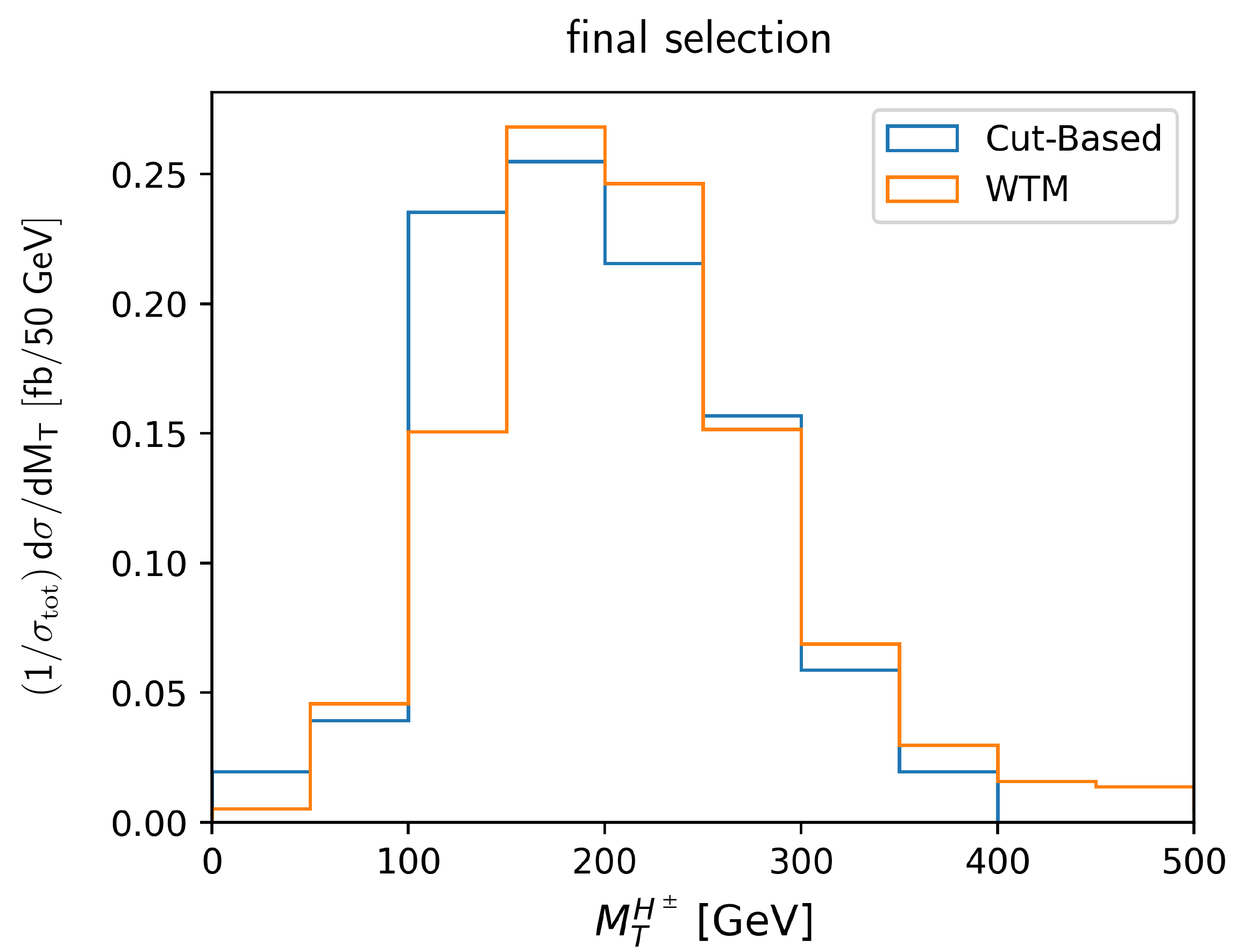}
\vspace{-0.2cm}
\caption{
Comparison of the $\mtch$ distribution from the $\wpm jj$ background in the cut-based analysis
with that in the WTM, after the final selection.}
\label{fig-cut-vs-WTM-mT-joint}
\end{figure}

Finally, to confirm the effectiveness of the WFM, we compare its resulting distributions with those from the traditional cut-based analysis. This comparison is meaningful when the cut-based analysis accurately reflects the main features of the true distribution after applying the final selection criteria.
However, the $m_{s_{1i}s_{2i}}$ distribution in the $\wpm jj$ background is not suitable for this comparison due to its complexity.
Therefore, we consider the $\mtch$ distribution for the comparison, which
enjoys a simple, smooth hill-like shape.

In \autoref{fig-cut-vs-WTM-mT-joint}, 
we present side-by-side the $\mtch$ distribution of the $\wpm jj$ background from the traditional cut-based analysis and its WFM counterpart.  
The results are post the final selection.
Despite the inherent constraints arising from the limited data in the cut-based method, 
there is a clear resemblance between the two distributions.
Both profiles exhibit a smoothly contoured hill shape and almost the same peak positions. 
This similarity underscores the capability of the WFM to properly represent the $\mtch$ distribution.
In conclusion, the WFM proves indispensable when tackling huge backgrounds with particularly stringent selection criteria.

\bibliographystyle{JHEP}  
\bibliography{ML-light-hf}  

\providecommand{\href}[2]{#2}\begingroup\raggedright\begin{thebibliography}{100}

\bibitem{ATLAS:2012yve}
{\scshape ATLAS} collaboration, G.~Aad et~al., \emph{{Observation of a new
  particle in the search for the Standard Model Higgs boson with the ATLAS
  detector at the LHC}},
  \href{http://dx.doi.org/10.1016/j.physletb.2012.08.020}{\emph{Phys. Lett. B}
  \textbf{ 716} (2012) 1--29},
  [\href{https://arxiv.org/abs/1207.7214}{\texttt{1207.7214}}].

\bibitem{CMS:2012qbp}
{\scshape CMS} collaboration, S.~Chatrchyan et~al., \emph{{Observation of a New
  Boson at a Mass of 125 GeV with the CMS Experiment at the LHC}},
  \href{http://dx.doi.org/10.1016/j.physletb.2012.08.021}{\emph{Phys. Lett. B}
  \textbf{ 716} (2012) 30--61},
  [\href{https://arxiv.org/abs/1207.7235}{\texttt{1207.7235}}].

\bibitem{Navarro:1995iw}
J.~F. Navarro, C.~S. Frenk and S.~D.~M. White, \emph{{The Structure of cold
  dark matter halos}}, \href{http://dx.doi.org/10.1086/177173}{\emph{Astrophys.
  J.} \textbf{ 462} (1996) 563--575},
  [\href{https://arxiv.org/abs/astro-ph/9508025}{\texttt{astro-ph/9508025}}].

\bibitem{Bertone:2004pz}
G.~Bertone, D.~Hooper and J.~Silk, \emph{{Particle dark matter: Evidence,
  candidates and constraints}},
  \href{http://dx.doi.org/10.1016/j.physrep.2004.08.031}{\emph{Phys. Rept.}
  \textbf{ 405} (2005) 279--390},
  [\href{https://arxiv.org/abs/hep-ph/0404175}{\texttt{hep-ph/0404175}}].

\bibitem{Degrassi:2012ry}
G.~Degrassi, S.~Di~Vita, J.~Elias-Miro, J.~R. Espinosa, G.~F. Giudice,
  G.~Isidori et~al., \emph{{Higgs mass and vacuum stability in the Standard
  Model at NNLO}}, \href{http://dx.doi.org/10.1007/JHEP08(2012)098}{\emph{JHEP}
  \textbf{ 08} (2012) 098},
  [\href{https://arxiv.org/abs/1205.6497}{\texttt{1205.6497}}].

\bibitem{Dimopoulos:1995mi}
S.~Dimopoulos and G.~F. Giudice, \emph{{Naturalness constraints in
  supersymmetric theories with nonuniversal soft terms}},
  \href{http://dx.doi.org/10.1016/0370-2693(95)00961-J}{\emph{Phys. Lett. B}
  \textbf{ 357} (1995) 573--578},
  [\href{https://arxiv.org/abs/hep-ph/9507282}{\texttt{hep-ph/9507282}}].

\bibitem{Chan:1997bi}
K.~L. Chan, U.~Chattopadhyay and P.~Nath, \emph{{Naturalness, weak scale
  supersymmetry and the prospect for the observation of supersymmetry at the
  Tevatron and at the CERN LHC}},
  \href{http://dx.doi.org/10.1103/PhysRevD.58.096004}{\emph{Phys. Rev. D}
  \textbf{ 58} (1998) 096004},
  [\href{https://arxiv.org/abs/hep-ph/9710473}{\texttt{hep-ph/9710473}}].

\bibitem{Craig:2015pha}
N.~Craig, A.~Katz, M.~Strassler and R.~Sundrum, \emph{{Naturalness in the Dark
  at the LHC}}, \href{http://dx.doi.org/10.1007/JHEP07(2015)105}{\emph{JHEP}
  \textbf{ 07} (2015) 105},
  [\href{https://arxiv.org/abs/1501.05310}{\texttt{1501.05310}}].

\bibitem{Akeroyd:1995hg}
A.~G. Akeroyd, \emph{{Fermiophobic Higgs bosons at the Tevatron}},
  \href{http://dx.doi.org/10.1016/0370-2693(95)01478-0}{\emph{Phys. Lett. B}
  \textbf{ 368} (1996) 89--95},
  [\href{https://arxiv.org/abs/hep-ph/9511347}{\texttt{hep-ph/9511347}}].

\bibitem{Akeroyd:1998ui}
A.~G. Akeroyd, \emph{{Fermiophobic and other nonminimal neutral Higgs bosons at
  the LHC}}, \href{http://dx.doi.org/10.1088/0954-3899/24/11/001}{\emph{J.
  Phys. G} \textbf{ 24} (1998) 1983--1994},
  [\href{https://arxiv.org/abs/hep-ph/9803324}{\texttt{hep-ph/9803324}}].

\bibitem{Akeroyd:1998dt}
A.~G. Akeroyd, \emph{{Three body decays of Higgs bosons at LEP-2 and
  application to a hidden fermiophobic Higgs}},
  \href{http://dx.doi.org/10.1016/S0550-3213(98)00845-1}{\emph{Nucl. Phys. B}
  \textbf{ 544} (1999) 557--575},
  [\href{https://arxiv.org/abs/hep-ph/9806337}{\texttt{hep-ph/9806337}}].

\bibitem{Barroso:1999bf}
A.~Barroso, L.~Brucher and R.~Santos, \emph{{Is there a light fermiophobic
  Higgs?}}, \href{http://dx.doi.org/10.1103/PhysRevD.60.035005}{\emph{Phys.
  Rev. D} \textbf{ 60} (1999) 035005},
  [\href{https://arxiv.org/abs/hep-ph/9901293}{\texttt{hep-ph/9901293}}].

\bibitem{Brucher:1999tx}
L.~Brucher and R.~Santos, \emph{{Experimental signatures of fermiophobic Higgs
  bosons}}, \href{http://dx.doi.org/10.1007/s100529900252}{\emph{Eur. Phys. J.
  C} \textbf{ 12} (2000) 87--98},
  [\href{https://arxiv.org/abs/hep-ph/9907434}{\texttt{hep-ph/9907434}}].

\bibitem{Akeroyd:2003bt}
A.~G. Akeroyd and M.~A. Diaz, \emph{{Searching for a light fermiophobic Higgs
  boson at the Tevatron}},
  \href{http://dx.doi.org/10.1103/PhysRevD.67.095007}{\emph{Phys. Rev. D}
  \textbf{ 67} (2003) 095007},
  [\href{https://arxiv.org/abs/hep-ph/0301203}{\texttt{hep-ph/0301203}}].

\bibitem{Akeroyd:2003xi}
A.~G. Akeroyd, M.~A. Diaz and F.~J. Pacheco, \emph{{Double fermiophobic Higgs
  boson production at the CERN LHC and LC}},
  \href{http://dx.doi.org/10.1103/PhysRevD.70.075002}{\emph{Phys. Rev. D}
  \textbf{ 70} (2004) 075002},
  [\href{https://arxiv.org/abs/hep-ph/0312231}{\texttt{hep-ph/0312231}}].

\bibitem{Akeroyd:2007yh}
A.~G. Akeroyd, M.~A. Diaz and M.~A. Rivera, \emph{{Effect of Charged Scalar
  Loops on Photonic Decays of a Fermiophobic Higgs}},
  \href{http://dx.doi.org/10.1103/PhysRevD.76.115012}{\emph{Phys. Rev. D}
  \textbf{ 76} (2007) 115012},
  [\href{https://arxiv.org/abs/0708.1939}{\texttt{0708.1939}}].

\bibitem{Arhrib:2008pw}
A.~Arhrib, R.~Benbrik, R.~B. Guedes and R.~Santos, \emph{{Search for a light
  fermiophobic Higgs boson produced via gluon fusion at Hadron Colliders}},
  \href{http://dx.doi.org/10.1103/PhysRevD.78.075002}{\emph{Phys. Rev. D}
  \textbf{ 78} (2008) 075002},
  [\href{https://arxiv.org/abs/0805.1603}{\texttt{0805.1603}}].

\bibitem{Gabrielli:2012yz}
E.~Gabrielli, B.~Mele and M.~Raidal, \emph{{Has a fermiophobic Higgs boson been
  detected at the LHC?}},
  \href{http://dx.doi.org/10.1016/j.physletb.2012.08.043}{\emph{Phys. Lett. B}
  \textbf{ 716} (2012) 322--325},
  [\href{https://arxiv.org/abs/1202.1796}{\texttt{1202.1796}}].

\bibitem{Berger:2012sy}
E.~L. Berger, Z.~Sullivan and H.~Zhang, \emph{{Associated Higgs plus vector
  boson test of a fermiophobic Higgs boson}},
  \href{http://dx.doi.org/10.1103/PhysRevD.86.015011}{\emph{Phys. Rev. D}
  \textbf{ 86} (2012) 015011},
  [\href{https://arxiv.org/abs/1203.6645}{\texttt{1203.6645}}].

\bibitem{Gabrielli:2012hd}
E.~Gabrielli, K.~Kannike, B.~Mele, A.~Racioppi and M.~Raidal,
  \emph{{Fermiophobic Higgs Boson and Supersymmetry}},
  \href{http://dx.doi.org/10.1103/PhysRevD.86.055014}{\emph{Phys. Rev. D}
  \textbf{ 86} (2012) 055014},
  [\href{https://arxiv.org/abs/1204.0080}{\texttt{1204.0080}}].

\bibitem{Cardenas:2012bg}
H.~Cardenas, A.~C.~B. Machado, V.~Pleitez and J.~A. Rodriguez, \emph{{Higgs
  decay rate to two photons in a model with two fermiophobic-Higgs doublets}},
  \href{http://dx.doi.org/10.1103/PhysRevD.87.035028}{\emph{Phys. Rev. D}
  \textbf{ 87} (2013) 035028},
  [\href{https://arxiv.org/abs/1212.1665}{\texttt{1212.1665}}].

\bibitem{Ilisie:2014hea}
V.~Ilisie and A.~Pich, \emph{{Low-mass fermiophobic charged Higgs phenomenology
  in two-Higgs-doublet models}},
  \href{http://dx.doi.org/10.1007/JHEP09(2014)089}{\emph{JHEP} \textbf{ 09}
  (2014) 089}, [\href{https://arxiv.org/abs/1405.6639}{\texttt{1405.6639}}].

\bibitem{Delgado:2016arn}
A.~Delgado, M.~Garcia-Pepin, M.~Quiros, J.~Santiago and R.~Vega-Morales,
  \emph{{Diphoton and Diboson Probes of Fermiophobic Higgs Bosons at the LHC}},
  \href{http://dx.doi.org/10.1007/JHEP06(2016)042}{\emph{JHEP} \textbf{ 06}
  (2016) 042}, [\href{https://arxiv.org/abs/1603.00962}{\texttt{1603.00962}}].

\bibitem{Mondal:2021bxa}
T.~Mondal and P.~Sanyal, \emph{{Same sign trilepton as signature of charged
  Higgs in two Higgs doublet model}},
  \href{http://dx.doi.org/10.1007/JHEP05(2022)040}{\emph{JHEP} \textbf{ 05}
  (2022) 040}, [\href{https://arxiv.org/abs/2109.05682}{\texttt{2109.05682}}].

\bibitem{Bahl:2021str}
H.~Bahl, T.~Stefaniak and J.~Wittbrodt, \emph{{The forgotten channels: charged
  Higgs boson decays to a W$^{±}$ and a non-SM-like Higgs boson}},
  \href{http://dx.doi.org/10.1007/JHEP06(2021)183}{\emph{JHEP} \textbf{ 06}
  (2021) 183}, [\href{https://arxiv.org/abs/2103.07484}{\texttt{2103.07484}}].

\bibitem{Kim:2022hvh}
J.~Kim, S.~Lee, P.~Sanyal and J.~Song, \emph{{CDF W-boson mass and muon g-2 in
  a type-X two-Higgs-doublet model with a Higgs-phobic light pseudoscalar}},
  \href{http://dx.doi.org/10.1103/PhysRevD.106.035002}{\emph{Phys. Rev. D}
  \textbf{ 106} (2022) 035002},
  [\href{https://arxiv.org/abs/2205.01701}{\texttt{2205.01701}}].

\bibitem{Kim:2023lxc}
J.~Kim, S.~Lee, P.~Sanyal, J.~Song and D.~Wang, \emph{{$\tau^{ \pm} \nu \gamma
  \gamma$ and $\ell^{ \pm} \ell^{ \pm} \gamma \gamma \rlap{\,/}{E}_T X$ to
  probe the fermiophobic Higgs boson with high cutoff scales}},
  \href{http://dx.doi.org/10.1007/JHEP04(2023)083}{\emph{JHEP} \textbf{ 04}
  (2023) 083}, [\href{https://arxiv.org/abs/2302.05467}{\texttt{2302.05467}}].

\bibitem{Mondal:2023wib}
T.~Mondal, S.~Moretti, S.~Munir and P.~Sanyal, \emph{{Electroweak multi-Higgs
  production: A smoking gun for the Type-I 2HDM}},
  \href{https://arxiv.org/abs/2304.07719}{\texttt{2304.07719}}.

\bibitem{Sanyal:2023pfs}
P.~Sanyal and D.~Wang, \emph{{Probing the electroweak $4b + \ell +
  {\rlap{\,/}{E}_T}$ final state in type I 2HDM at the LHC}},
  \href{http://dx.doi.org/10.1007/JHEP09(2023)076}{\emph{JHEP} \textbf{ 09}
  (2023) 076}, [\href{https://arxiv.org/abs/2305.00659}{\texttt{2305.00659}}].

\bibitem{Bernon:2015wef}
J.~Bernon, J.~F. Gunion, H.~E. Haber, Y.~Jiang and S.~Kraml,
  \emph{{Scrutinizing the alignment limit in two-Higgs-doublet models. II.
  m$_H$=125 GeV}},
  \href{http://dx.doi.org/10.1103/PhysRevD.93.035027}{\emph{Phys. Rev. D}
  \textbf{ 93} (2016) 035027},
  [\href{https://arxiv.org/abs/1511.03682}{\texttt{1511.03682}}].

\bibitem{Chang:2015goa}
S.~Chang, S.~K. Kang, J.-P. Lee and J.~Song, \emph{{Higgs potential and hidden
  light Higgs scenario in two Higgs doublet models}},
  \href{http://dx.doi.org/10.1103/PhysRevD.92.075023}{\emph{Phys. Rev. D}
  \textbf{ 92} (2015) 075023},
  [\href{https://arxiv.org/abs/1507.03618}{\texttt{1507.03618}}].

\bibitem{Song:2019aav}
J.~Song and Y.~W. Yoon, \emph{{$W\gamma$ decay of the elusive charged Higgs
  boson in the two-Higgs-doublet model with vectorlike fermions}},
  \href{http://dx.doi.org/10.1103/PhysRevD.100.055006}{\emph{Phys. Rev. D}
  \textbf{ 100} (2019) 055006},
  [\href{https://arxiv.org/abs/1904.06521}{\texttt{1904.06521}}].

\bibitem{Jueid:2021avn}
A.~Jueid, J.~Kim, S.~Lee and J.~Song, \emph{{Type-X two-Higgs-doublet model in
  light of the muon g-2: Confronting Higgs boson and collider data}},
  \href{http://dx.doi.org/10.1103/PhysRevD.104.095008}{\emph{Phys. Rev. D}
  \textbf{ 104} (2021) 095008},
  [\href{https://arxiv.org/abs/2104.10175}{\texttt{2104.10175}}].

\bibitem{Cheung:2022ndq}
K.~Cheung, A.~Jueid, J.~Kim, S.~Lee, C.-T. Lu and J.~Song, \emph{{Comprehensive
  study of the light charged Higgs boson in the type-I two-Higgs-doublet
  model}}, \href{http://dx.doi.org/10.1103/PhysRevD.105.095044}{\emph{Phys.
  Rev. D} \textbf{ 105} (2022) 095044},
  [\href{https://arxiv.org/abs/2201.06890}{\texttt{2201.06890}}].

\bibitem{Lee:2022gyf}
S.~Lee, K.~Cheung, J.~Kim, C.-T. Lu and J.~Song, \emph{{Status of the
  two-Higgs-doublet model in light of the CDF mW measurement}},
  \href{http://dx.doi.org/10.1103/PhysRevD.106.075013}{\emph{Phys. Rev. D}
  \textbf{ 106} (2022) 075013},
  [\href{https://arxiv.org/abs/2204.10338}{\texttt{2204.10338}}].

\bibitem{Akeroyd:2005pr}
A.~G. Akeroyd, A.~Alves, M.~A. Diaz and O.~J.~P. Eboli, \emph{{Multi-photon
  signatures at the Fermilab Tevatron}},
  \href{http://dx.doi.org/10.1140/epjc/s10052-006-0014-8}{\emph{Eur. Phys. J.
  C} \textbf{ 48} (2006) 147--157},
  [\href{https://arxiv.org/abs/hep-ph/0512077}{\texttt{hep-ph/0512077}}].

\bibitem{Arhrib:2017wmo}
A.~Arhrib, R.~Benbrik, R.~Enberg, W.~Klemm, S.~Moretti and S.~Munir,
  \emph{{Identifying a light charged Higgs boson at the LHC Run II}},
  \href{http://dx.doi.org/10.1016/j.physletb.2017.10.006}{\emph{Phys. Lett. B}
  \textbf{ 774} (2017) 591--598},
  [\href{https://arxiv.org/abs/1706.01964}{\texttt{1706.01964}}].

\bibitem{Kim:2022nmm}
J.~Kim, S.~Lee, J.~Song and P.~Sanyal, \emph{{Fermiophobic light Higgs boson in
  the type-I two-Higgs-doublet model}},
  \href{http://dx.doi.org/10.1016/j.physletb.2022.137406}{\emph{Phys. Lett. B}
  \textbf{ 834} (2022) 137406},
  [\href{https://arxiv.org/abs/2207.05104}{\texttt{2207.05104}}].

\bibitem{deFavereau:2013fsa}
{\scshape DELPHES 3} collaboration, J.~de~Favereau, C.~Delaere, P.~Demin,
  A.~Giammanco, V.~Lema\^\i{}tre, A.~Mertens et~al., \emph{{DELPHES 3, A
  modular framework for fast simulation of a generic collider experiment}},
  \href{http://dx.doi.org/10.1007/JHEP02(2014)057}{\emph{JHEP} \textbf{ 02}
  (2014) 057}, [\href{https://arxiv.org/abs/1307.6346}{\texttt{1307.6346}}].

\bibitem{Mimasu:2014nea}
K.~Mimasu and V.~Sanz, \emph{{ALPs at Colliders}},
  \href{http://dx.doi.org/10.1007/JHEP06(2015)173}{\emph{JHEP} \textbf{ 06}
  (2015) 173}, [\href{https://arxiv.org/abs/1409.4792}{\texttt{1409.4792}}].

\bibitem{Bauer:2017nlg}
M.~Bauer, M.~Neubert and A.~Thamm, \emph{{LHC as an Axion Factory: Probing an
  Axion Explanation for $(g-2)_\mu$ with Exotic Higgs Decays}},
  \href{http://dx.doi.org/10.1103/PhysRevLett.119.031802}{\emph{Phys. Rev.
  Lett.} \textbf{ 119} (2017) 031802},
  [\href{https://arxiv.org/abs/1704.08207}{\texttt{1704.08207}}].

\bibitem{Knapen:2021elo}
S.~Knapen, S.~Kumar and D.~Redigolo, \emph{{Searching for axionlike particles
  with data scouting at ATLAS and CMS}},
  \href{http://dx.doi.org/10.1103/PhysRevD.105.115012}{\emph{Phys. Rev. D}
  \textbf{ 105} (2022) 115012},
  [\href{https://arxiv.org/abs/2112.07720}{\texttt{2112.07720}}].

\bibitem{Wang:2021uyb}
D.~Wang, L.~Wu, J.~M. Yang and M.~Zhang, \emph{{Photon-jet events as a probe of
  axionlike particles at the LHC}},
  \href{http://dx.doi.org/10.1103/PhysRevD.104.095016}{\emph{Phys. Rev. D}
  \textbf{ 104} (2021) 095016},
  [\href{https://arxiv.org/abs/2102.01532}{\texttt{2102.01532}}].

\bibitem{Ren:2021prq}
J.~Ren, D.~Wang, L.~Wu, J.~M. Yang and M.~Zhang, \emph{{Detecting an axion-like
  particle with machine learning at the LHC}},
  \href{http://dx.doi.org/10.1007/JHEP11(2021)138}{\emph{JHEP} \textbf{ 11}
  (2021) 138}, [\href{https://arxiv.org/abs/2106.07018}{\texttt{2106.07018}}].

\bibitem{Larkoski:2017jix}
A.~J. Larkoski, I.~Moult and B.~Nachman, \emph{{Jet Substructure at the Large
  Hadron Collider: A Review of Recent Advances in Theory and Machine
  Learning}},
  \href{http://dx.doi.org/10.1016/j.physrep.2019.11.001}{\emph{Phys. Rept.}
  \textbf{ 841} (2020) 1--63},
  [\href{https://arxiv.org/abs/1709.04464}{\texttt{1709.04464}}].

\bibitem{Guest:2018yhq}
D.~Guest, K.~Cranmer and D.~Whiteson, \emph{{Deep Learning and its Application
  to LHC Physics}},
  \href{http://dx.doi.org/10.1146/annurev-nucl-101917-021019}{\emph{Ann. Rev.
  Nucl. Part. Sci.} \textbf{ 68} (2018) 161--181},
  [\href{https://arxiv.org/abs/1806.11484}{\texttt{1806.11484}}].

\bibitem{Albertsson:2018maf}
K.~Albertsson et~al., \emph{{Machine Learning in High Energy Physics Community
  White Paper}},
  \href{http://dx.doi.org/10.1088/1742-6596/1085/2/022008}{\emph{J. Phys. Conf.
  Ser.} \textbf{ 1085} (2018) 022008},
  [\href{https://arxiv.org/abs/1807.02876}{\texttt{1807.02876}}].

\bibitem{Radovic:2018dip}
A.~Radovic, M.~Williams, D.~Rousseau, M.~Kagan, D.~Bonacorsi, A.~Himmel et~al.,
  \emph{{Machine learning at the energy and intensity frontiers of particle
  physics}}, \href{http://dx.doi.org/10.1038/s41586-018-0361-2}{\emph{Nature}
  \textbf{ 560} (2018) 41--48}.

\bibitem{Bourilkov:2019yoi}
D.~Bourilkov, \emph{{Machine and Deep Learning Applications in Particle
  Physics}}, \href{http://dx.doi.org/10.1142/S0217751X19300199}{\emph{Int. J.
  Mod. Phys. A} \textbf{ 34} (2020) 1930019},
  [\href{https://arxiv.org/abs/1912.08245}{\texttt{1912.08245}}].

\bibitem{Feickert:2021ajf}
M.~Feickert and B.~Nachman, \emph{{A Living Review of Machine Learning for
  Particle Physics}},
  \href{https://arxiv.org/abs/2102.02770}{\texttt{2102.02770}}.

\bibitem{Shanahan:2022ifi}
P.~Shanahan et~al., \emph{{Snowmass 2021 Computational Frontier CompF03 Topical
  Group Report: Machine Learning}},
  \href{https://arxiv.org/abs/2209.07559}{\texttt{2209.07559}}.

\bibitem{CMS:2019gpd}
{\scshape CMS} collaboration, \emph{{Machine learning-based identification of
  highly Lorentz-boosted hadronically decaying particles at the CMS
  experiment}},   \href{http://cds.cern.ch/record/2683870}{\textit{CMS-PAS-JME-18-002}}.

\bibitem{Branco:2011iw}
G.~C. Branco, P.~M. Ferreira, L.~Lavoura, M.~N. Rebelo, M.~Sher and J.~P.
  Silva, \emph{{Theory and phenomenology of two-Higgs-doublet models}},
  \href{http://dx.doi.org/10.1016/j.physrep.2012.02.002}{\emph{Phys. Rept.}
  \textbf{ 516} (2012) 1--102},
  [\href{https://arxiv.org/abs/1106.0034}{\texttt{1106.0034}}].

\bibitem{Glashow:1976nt}
S.~L. Glashow and S.~Weinberg, \emph{{Natural Conservation Laws for Neutral
  Currents}}, \href{http://dx.doi.org/10.1103/PhysRevD.15.1958}{\emph{Phys.
  Rev. D} \textbf{ 15} (1977) 1958}.

\bibitem{Paschos:1976ay}
E.~A. Paschos, \emph{{Diagonal Neutral Currents}},
  \href{http://dx.doi.org/10.1103/PhysRevD.15.1966}{\emph{Phys. Rev. D}
  \textbf{ 15} (1977) 1966}.

\bibitem{ATLAS:2020fcp}
{\scshape ATLAS} collaboration, G.~Aad et~al., \emph{{Measurements of $WH$ and
  $ZH$ production in the $H \rightarrow b\bar{b}$ decay channel in $pp$
  collisions at 13 TeV with the ATLAS detector}},
  \href{http://dx.doi.org/10.1140/epjc/s10052-020-08677-2}{\emph{Eur. Phys. J.
  C} \textbf{ 81} (2021) 178},
  [\href{https://arxiv.org/abs/2007.02873}{\texttt{2007.02873}}].

\bibitem{ATLAS:2020bhl}
{\scshape ATLAS} collaboration, G.~Aad et~al., \emph{{Measurements of Higgs
  bosons decaying to bottom quarks from vector boson fusion production with the
  ATLAS experiment at $\sqrt{s}=13\,\text {TeV}$}},
  \href{http://dx.doi.org/10.1140/epjc/s10052-021-09192-8}{\emph{Eur. Phys. J.
  C} \textbf{ 81} (2021) 537},
  [\href{https://arxiv.org/abs/2011.08280}{\texttt{2011.08280}}].

\bibitem{CMS:2020zge}
{\scshape CMS} collaboration, A.~M. Sirunyan et~al., \emph{{Inclusive search
  for highly boosted Higgs bosons decaying to bottom quark-antiquark pairs in
  proton-proton collisions at $\sqrt{s} =$ 13 TeV}},
  \href{http://dx.doi.org/10.1007/JHEP12(2020)085}{\emph{JHEP} \textbf{ 12}
  (2020) 085}, [\href{https://arxiv.org/abs/2006.13251}{\texttt{2006.13251}}].

\bibitem{ATLAS:2021nsx}
{\scshape ATLAS} collaboration, \emph{{Study of Higgs-boson production with
  large transverse momentum using the $H\rightarrow b\bar{b}$ decay with the
  ATLAS detector}}, \href{http://cds.cern.ch/record/2759284}{\textit{ATLAS-CONF-2021-010}}.

\bibitem{CMS:2021gxc}
{\scshape CMS} collaboration, A.~Tumasyan et~al., \emph{{Measurement of the
  inclusive and differential Higgs boson production cross sections in the decay
  mode to a pair of $\tau$ leptons in pp collisions at $\sqrt{s} = $ 13 TeV}},
  \href{http://dx.doi.org/10.1103/PhysRevLett.128.081805}{\emph{Phys. Rev.
  Lett.} \textbf{ 128} (2022) 081805},
  [\href{https://arxiv.org/abs/2107.11486}{\texttt{2107.11486}}].

\bibitem{ATLAS:2020syy}
{\scshape ATLAS} collaboration, \emph{{Measurement of the Higgs boson decaying
  to $b$-quarks produced in association with a top-quark pair in $pp$
  collisions at $\sqrt{s}=13$ TeV with the ATLAS detector}}, \href{http://cds.cern.ch/record/2743685}{\textit{ATLAS-CONF-2020-058}}.

\bibitem{ATLAS:2021upe}
{\scshape ATLAS} collaboration, \emph{{Measurements of gluon fusion and
  vector-boson-fusion production of the Higgs boson in $H\rightarrow W W^*
  \rightarrow e\nu \mu\nu$ decays using $pp$ collisions at $\sqrt{s}=13$ TeV
  with the ATLAS detector}}, \href{http://cds.cern.ch/record/2759651}{\textit{ATLAS-CONF-2021-014}}.


\bibitem{ATLAS:2020pvn}
{\scshape ATLAS} collaboration, \emph{{Measurement of the properties of Higgs
  boson production at $\sqrt{s}$=13 TeV in the $H\to \gamma\gamma$ channel
  using 139 fb$^{-1}$ of $pp$ collision data with the ATLAS experiment}}, 
   \href{http://cds.cern.ch/record/2725727}{\textit{ATLAS-CONF-2020-026}}.

\bibitem{CMS:2021ugl}
{\scshape CMS} collaboration, A.~M. Sirunyan et~al., \emph{{Measurements of
  production cross sections of the Higgs boson in the four-lepton final state
  in proton\textendash{}proton collisions at $\sqrt{s} = 13\,\text {Te}\text
  {V} $}}, \href{http://dx.doi.org/10.1140/epjc/s10052-021-09200-x}{\emph{Eur.
  Phys. J. C} \textbf{ 81} (2021) 488},
  [\href{https://arxiv.org/abs/2103.04956}{\texttt{2103.04956}}].

\bibitem{ATLAS:2020wny}
{\scshape ATLAS} collaboration, G.~Aad et~al., \emph{{Measurements of the Higgs
  boson inclusive and differential fiducial cross sections in the 4$\ell $
  decay channel at $\sqrt{s}$ = 13 TeV}},
  \href{http://dx.doi.org/10.1140/epjc/s10052-020-8223-0}{\emph{Eur. Phys. J.
  C} \textbf{ 80} (2020) 942},
  [\href{https://arxiv.org/abs/2004.03969}{\texttt{2004.03969}}].

\bibitem{ATLAS:2020rej}
{\scshape ATLAS} collaboration, G.~Aad et~al., \emph{{Higgs boson production
  cross-section measurements and their EFT interpretation in the $4\ell $ decay
  channel at $\sqrt{s}=$13 TeV with the ATLAS detector}},
  \href{http://dx.doi.org/10.1140/epjc/s10052-020-8227-9}{\emph{Eur. Phys. J.
  C} \textbf{ 80} (2020) 957},
  [\href{https://arxiv.org/abs/2004.03447}{\texttt{2004.03447}}].

\bibitem{ATLAS:2020qdt}
{\scshape ATLAS} collaboration, \emph{{A combination of measurements of Higgs
  boson production and decay using up to $139$ fb$^{-1}$ of proton--proton
  collision data at $\sqrt{s}=$ 13 TeV collected with the ATLAS experiment}}, \href{http://cds.cern.ch/record/2725733}{\textit{ATLAS-CONF-2020-027}}.

\bibitem{ATLAS:2020fzp}
{\scshape ATLAS} collaboration, G.~Aad et~al., \emph{{A search for the dimuon
  decay of the Standard Model Higgs boson with the ATLAS detector}},
  \href{http://dx.doi.org/10.1016/j.physletb.2020.135980}{\emph{Phys. Lett. B}
  \textbf{ 812} (2021) 135980},
  [\href{https://arxiv.org/abs/2007.07830}{\texttt{2007.07830}}].

\bibitem{CMS:2020xwi}
{\scshape CMS} collaboration, A.~M. Sirunyan et~al., \emph{{Evidence for Higgs
  boson decay to a pair of muons}},
  \href{http://dx.doi.org/10.1007/JHEP01(2021)148}{\emph{JHEP} \textbf{ 01}
  (2021) 148}, [\href{https://arxiv.org/abs/2009.04363}{\texttt{2009.04363}}].

\bibitem{ATLAS:2021zwx}
{\scshape ATLAS} collaboration, \emph{{Direct constraint on the Higgs-charm
  coupling from a search for Higgs boson decays to charm quarks with the ATLAS
  detector}}, \href{http://cds.cern.ch/record/2771724}{\textit{ATLAS-CONF-2021-021}}.

\bibitem{Peskin:1991sw}
M.~E. Peskin and T.~Takeuchi, \emph{{Estimation of oblique electroweak
  corrections}}, \href{http://dx.doi.org/10.1103/PhysRevD.46.381}{\emph{Phys.
  Rev. D} \textbf{ 46} (1992) 381--409}.

\bibitem{He:2001tp}
H.-J. He, N.~Polonsky and S.-f. Su, \emph{{Extra families, Higgs spectrum and
  oblique corrections}},
  \href{http://dx.doi.org/10.1103/PhysRevD.64.053004}{\emph{Phys. Rev. D}
  \textbf{ 64} (2001) 053004},
  [\href{https://arxiv.org/abs/hep-ph/0102144}{\texttt{hep-ph/0102144}}].

\bibitem{Grimus:2008nb}
W.~Grimus, L.~Lavoura, O.~M. Ogreid and P.~Osland, \emph{{The Oblique
  parameters in multi-Higgs-doublet models}},
  \href{http://dx.doi.org/10.1016/j.nuclphysb.2008.04.019}{\emph{Nucl. Phys. B}
  \textbf{ 801} (2008) 81--96},
  [\href{https://arxiv.org/abs/0802.4353}{\texttt{0802.4353}}].

\bibitem{Ivanov:2006yq}
I.~P. Ivanov, \emph{{Minkowski space structure of the Higgs potential in
  2HDM}}, \href{http://dx.doi.org/10.1103/PhysRevD.75.035001}{\emph{Phys. Rev.
  D} \textbf{ 75} (2007) 035001},
  [\href{https://arxiv.org/abs/hep-ph/0609018}{\texttt{hep-ph/0609018}}].

\bibitem{Arhrib:2000is}
A.~Arhrib, \emph{{Unitarity constraints on scalar parameters of the standard
  and two Higgs doublets model}},  in \emph{{Workshop on Noncommutative
  Geometry, Superstrings and Particle Physics}}, 12, 2000.
\newblock \href{https://arxiv.org/abs/hep-ph/0012353}{\texttt{hep-ph/0012353}}.

\bibitem{Ivanov:2008cxa}
I.~P. Ivanov, \emph{{General two-order-parameter Ginzburg-Landau model with
  quadratic and quartic interactions}},
  \href{http://dx.doi.org/10.1103/PhysRevE.79.021116}{\emph{Phys. Rev. E}
  \textbf{ 79} (2009) 021116},
  [\href{https://arxiv.org/abs/0802.2107}{\texttt{0802.2107}}].

\bibitem{Barroso:2012mj}
A.~Barroso, P.~M. Ferreira, I.~P. Ivanov, R.~Santos and J.~P. Silva,
  \emph{{Evading death by vacuum}},
  \href{http://dx.doi.org/10.1140/epjc/s10052-013-2537-0}{\emph{Eur. Phys. J.
  C} \textbf{ 73} (2013) 2537},
  [\href{https://arxiv.org/abs/1211.6119}{\texttt{1211.6119}}].

\bibitem{Barroso:2013awa}
A.~Barroso, P.~M. Ferreira, I.~P. Ivanov and R.~Santos, \emph{{Metastability
  bounds on the two Higgs doublet model}},
  \href{http://dx.doi.org/10.1007/JHEP06(2013)045}{\emph{JHEP} \textbf{ 06}
  (2013) 045}, [\href{https://arxiv.org/abs/1303.5098}{\texttt{1303.5098}}].

\bibitem{Arbey:2017gmh}
A.~Arbey, F.~Mahmoudi, O.~Stal and T.~Stefaniak, \emph{{Status of the Charged
  Higgs Boson in Two Higgs Doublet Models}},
  \href{http://dx.doi.org/10.1140/epjc/s10052-018-5651-1}{\emph{Eur. Phys. J.
  C} \textbf{ 78} (2018) 182},
  [\href{https://arxiv.org/abs/1706.07414}{\texttt{1706.07414}}].

\bibitem{Sanyal:2019xcp}
P.~Sanyal, \emph{{Limits on the Charged Higgs Parameters in the Two Higgs
  Doublet Model using CMS $\sqrt{s}=13$ TeV Results}},
  \href{http://dx.doi.org/10.1140/epjc/s10052-019-7431-y}{\emph{Eur. Phys. J.
  C} \textbf{ 79} (2019) 913},
  [\href{https://arxiv.org/abs/1906.02520}{\texttt{1906.02520}}].

\bibitem{Misiak:2017bgg}
M.~Misiak and M.~Steinhauser, \emph{{Weak radiative decays of the B meson and
  bounds on $M_{H^\pm }$ in the Two-Higgs-Doublet Model}},
  \href{http://dx.doi.org/10.1140/epjc/s10052-017-4776-y}{\emph{Eur. Phys. J.
  C} \textbf{ 77} (2017) 201},
  [\href{https://arxiv.org/abs/1702.04571}{\texttt{1702.04571}}].

\bibitem{Oredsson:2018yho}
J.~Oredsson and J.~Rathsman, \emph{{$\mathbb Z_2$ breaking effects in 2-loop RG
  evolution of 2HDM}},
  \href{http://dx.doi.org/10.1007/JHEP02(2019)152}{\emph{JHEP} \textbf{ 02}
  (2019) 152}, [\href{https://arxiv.org/abs/1810.02588}{\texttt{1810.02588}}].

\bibitem{Oredsson:2018vio}
J.~Oredsson, \emph{{2HDME : Two-Higgs-Doublet Model Evolver}},
  \href{http://dx.doi.org/10.1016/j.cpc.2019.05.021}{\emph{Comput. Phys.
  Commun.} \textbf{ 244} (2019) 409--426},
  [\href{https://arxiv.org/abs/1811.08215}{\texttt{1811.08215}}].

\bibitem{Bechtle:2013wla}
P.~Bechtle, O.~Brein, S.~Heinemeyer, O.~St\r{a}l, T.~Stefaniak, G.~Weiglein
  et~al., \emph{{$\mathsf{HiggsBounds}-4$: Improved Tests of Extended Higgs
  Sectors against Exclusion Bounds from LEP, the Tevatron and the LHC}},
  \href{http://dx.doi.org/10.1140/epjc/s10052-013-2693-2}{\emph{Eur. Phys. J.
  C} \textbf{ 74} (2014) 2693},
  [\href{https://arxiv.org/abs/1311.0055}{\texttt{1311.0055}}].

\bibitem{Bechtle:2020uwn}
P.~Bechtle, S.~Heinemeyer, T.~Klingl, T.~Stefaniak, G.~Weiglein and
  J.~Wittbrodt, \emph{{HiggsSignals-2: Probing new physics with precision Higgs
  measurements in the LHC 13 TeV era}},
  \href{http://dx.doi.org/10.1140/epjc/s10052-021-08942-y}{\emph{Eur. Phys. J.
  C} \textbf{ 81} (2021) 145},
  [\href{https://arxiv.org/abs/2012.09197}{\texttt{2012.09197}}].

\bibitem{DELPHI:2003hpv}
{\scshape DELPHI} collaboration, J.~Abdallah et~al., \emph{{Search for
  fermiophobic Higgs bosons in final states with photons at LEP 2}},
  \href{http://dx.doi.org/10.1140/epjc/s2004-01869-2}{\emph{Eur. Phys. J. C}
  \textbf{ 35} (2004) 313--324},
  [\href{https://arxiv.org/abs/hep-ex/0406012}{\texttt{hep-ex/0406012}}].

\bibitem{CDF:2016ybe}
{\scshape CDF} collaboration, T.~A. Aaltonen et~al., \emph{{Search for a
  Low-Mass Neutral Higgs Boson with Suppressed Couplings to Fermions Using
  Events with Multiphoton Final States}},
  \href{http://dx.doi.org/10.1103/PhysRevD.93.112010}{\emph{Phys. Rev. D}
  \textbf{ 93} (2016) 112010},
  [\href{https://arxiv.org/abs/1601.00401}{\texttt{1601.00401}}].

\bibitem{CMS:2021bvh}
{\scshape CMS} collaboration, \emph{{Search for exotic decay of the Higgs boson
  into two light pseudoscalars with four photons in the final state at
  $\sqrt{s}$ = 13 TeV}}, \href{http://cds.cern.ch/record/2776775}{\textit{CMS-PAS-HIG-21-003}}.

\bibitem{Kang:2022mdy}
S.~K. Kang, J.~Kim, S.~Lee and J.~Song, \emph{{Disentangling the high- and
  low-cutoff scales via the trilinear Higgs couplings in the type-I
  two-Higgs-doublet model}},
  \href{http://dx.doi.org/10.1103/PhysRevD.107.015025}{\emph{Phys. Rev. D}
  \textbf{ 107} (2023) 015025},
  [\href{https://arxiv.org/abs/2210.00020}{\texttt{2210.00020}}].

\bibitem{Aad:2015bua}
{\scshape ATLAS} collaboration, G.~Aad et~al., \emph{{Search for new phenomena
  in events with at least three photons collected in $pp$ collisions at
  $\sqrt{s}$ = 8 TeV with the ATLAS detector}},
  \href{http://dx.doi.org/10.1140/epjc/s10052-016-4034-8}{\emph{Eur. Phys. J.
  C} \textbf{ 76} (2016) 210},
  [\href{https://arxiv.org/abs/1509.05051}{\texttt{1509.05051}}].

\bibitem{ALEPH:2002gcw}
\emph{{Searches for Higgs Bosons Decaying into Photons : Combined Results from
  the LEP Experiments}}, \href{http://cds.cern.ch/record/644922}{\textit{CERN-ALEPH-2002-019}}.


\bibitem{Degrande:2011ua}
C.~Degrande, C.~Duhr, B.~Fuks, D.~Grellscheid, O.~Mattelaer and T.~Reiter,
  \emph{{UFO - The Universal FeynRules Output}},
  \href{http://dx.doi.org/10.1016/j.cpc.2012.01.022}{\emph{Comput. Phys.
  Commun.} \textbf{ 183} (2012) 1201--1214},
  [\href{https://arxiv.org/abs/1108.2040}{\texttt{1108.2040}}].

\bibitem{Alloul:2013bka}
A.~Alloul, N.~D. Christensen, C.~Degrande, C.~Duhr and B.~Fuks,
  \emph{{FeynRules 2.0 - A complete toolbox for tree-level phenomenology}},
  \href{http://dx.doi.org/10.1016/j.cpc.2014.04.012}{\emph{Comput. Phys.
  Commun.} \textbf{ 185} (2014) 2250--2300},
  [\href{https://arxiv.org/abs/1310.1921}{\texttt{1310.1921}}].

\bibitem{Alwall:2011uj}
J.~Alwall, M.~Herquet, F.~Maltoni, O.~Mattelaer and T.~Stelzer, \emph{{MadGraph
  5 : Going Beyond}},
  \href{http://dx.doi.org/10.1007/JHEP06(2011)128}{\emph{JHEP} \textbf{ 06}
  (2011) 128}, [\href{https://arxiv.org/abs/1106.0522}{\texttt{1106.0522}}].

\bibitem{NNPDF:2017mvq}
{\scshape NNPDF} collaboration, R.~D. Ball et~al., \emph{{Parton distributions
  from high-precision collider data}},
  \href{http://dx.doi.org/10.1140/epjc/s10052-017-5199-5}{\emph{Eur. Phys. J.
  C} \textbf{ 77} (2017) 663},
  [\href{https://arxiv.org/abs/1706.00428}{\texttt{1706.00428}}].

\bibitem{Eriksson:2009ws}
D.~Eriksson, J.~Rathsman and O.~Stal, \emph{{2HDMC: Two-Higgs-Doublet Model
  Calculator Physics and Manual}},
  \href{http://dx.doi.org/10.1016/j.cpc.2009.09.011}{\emph{Comput. Phys.
  Commun.} \textbf{ 181} (2010) 189--205},
  [\href{https://arxiv.org/abs/0902.0851}{\texttt{0902.0851}}].


\bibitem{Bierlich:2022pfr}
C.~Bierlich et~al., \emph{{A comprehensive guide to the physics and usage of
  PYTHIA 8.3}},  \href{https://arxiv.org/abs/2203.11601}{\texttt{2203.11601}}.

\bibitem{Cacciari:2011ma}
M.~Cacciari, G.~P. Salam and G.~Soyez, \emph{{FastJet User Manual}},
  \href{http://dx.doi.org/10.1140/epjc/s10052-012-1896-2}{\emph{Eur. Phys. J.
  C} \textbf{ 72} (2012) 1896},
  [\href{https://arxiv.org/abs/1111.6097}{\texttt{1111.6097}}].
  
  \bibitem{Cacciari:2008gp}
M.~Cacciari, G.~P. Salam and G.~Soyez, \emph{{The anti-$k_t$ jet clustering
  algorithm}},
  \href{http://dx.doi.org/10.1088/1126-6708/2008/04/063}{\emph{JHEP} \textbf{
  04} (2008) 063},
  [\href{https://arxiv.org/abs/0802.1189}{\texttt{0802.1189}}].


\bibitem{TheATLAScollaboration:2013pia}
{\scshape ATLAS} collaboration, \emph{{Pile-up subtraction and suppression for
  jets in ATLAS}}, 
\href{http://cds.cern.ch/record/1570994}{\textit{ATLAS-CONF-2013-083}}.


\bibitem{CMS:2014ata}
{\scshape CMS} collaboration, \emph{{Pileup Removal Algorithms}}, 
\href{http://cds.cern.ch/record/1751454}{\textit{CMS-PAS-JME-14-001}}.


\bibitem{Kirschenmann:2014dla}
{\scshape CMS} collaboration, H.~Kirschenmann, \emph{{Jet performance in CMS}},
  \href{http://dx.doi.org/10.22323/1.180.0433}{\emph{PoS} \textbf{ EPS-HEP2013}
  (2013) 433}.

\bibitem{Bertolini:2014bba}
D.~Bertolini, P.~Harris, M.~Low and N.~Tran, \emph{{Pileup Per Particle
  Identification}},
  \href{http://dx.doi.org/10.1007/JHEP10(2014)059}{\emph{JHEP} \textbf{ 10}
  (2014) 059}, [\href{https://arxiv.org/abs/1407.6013}{\texttt{1407.6013}}].

\bibitem{Cacciari:2014gra}
M.~Cacciari, G.~P. Salam and G.~Soyez, \emph{{SoftKiller, a particle-level
  pileup removal method}},
  \href{http://dx.doi.org/10.1140/epjc/s10052-015-3267-2}{\emph{Eur. Phys. J.
  C} \textbf{ 75} (2015) 59},
  [\href{https://arxiv.org/abs/1407.0408}{\texttt{1407.0408}}].

\bibitem{Catani:1993hr}
S.~Catani, Y.~L. Dokshitzer, M.~H. Seymour and B.~R. Webber,
  \emph{{Longitudinally invariant $K_t$ clustering algorithms for hadron hadron
  collisions}},
  \href{http://dx.doi.org/10.1016/0550-3213(93)90166-M}{\emph{Nucl. Phys. B}
  \textbf{ 406} (1993) 187--224}.

\bibitem{Ellis:1993tq}
S.~D. Ellis and D.~E. Soper, \emph{{Successive combination jet algorithm for
  hadron collisions}},
  \href{http://dx.doi.org/10.1103/PhysRevD.48.3160}{\emph{Phys. Rev. D}
  \textbf{ 48} (1993) 3160--3166},
  [\href{https://arxiv.org/abs/hep-ph/9305266}{\texttt{hep-ph/9305266}}].

\bibitem{Cowan:2010js}
G.~Cowan, K.~Cranmer, E.~Gross and O.~Vitells, \emph{{Asymptotic formulae for
  likelihood-based tests of new physics}},
  \href{http://dx.doi.org/10.1140/epjc/s10052-011-1554-0}{\emph{Eur. Phys. J.
  C} \textbf{ 71} (2011) 1554},
  [\href{https://arxiv.org/abs/1007.1727}{\texttt{1007.1727}}].

\bibitem{paszke2019pytorch}
A.~Paszke, S.~Gross, F.~Massa, A.~Lerer, J.~Bradbury, G.~Chanan et~al.,
  \emph{Pytorch: An imperative style, high-performance deep learning library},
  {\emph{Advances in neural information processing systems} \textbf{ 32} (2019)
  }, [\href{https://arxiv.org/abs/1912.01703}{\texttt{1912.01703}}].


\bibitem{loshchilov2017decoupled}
I.~Loshchilov and F.~Hutter, \emph{Decoupled weight decay regularization},
 [\href{https://arxiv.org/abs/1711.05101}{\texttt{1711.05101}}].

\end{thebibliography}\endgroup



\end{document}